\documentclass[amsmath,amssymb,twocolumn,showpacs,aps,prb,showkeys,longbibliography, 10pt]{revtex4-2}

\usepackage{graphicx}
\usepackage{amssymb}
\usepackage{ifthen}
\usepackage{color}
\usepackage{bm}
\usepackage{braket}
\usepackage{multirow}
\usepackage{soul}
\usepackage{tabularx}
\usepackage{booktabs}
\usepackage{array}
\usepackage{adjustbox}
\usepackage{siunitx}
\usepackage[table]{xcolor}
\usepackage{hyperref}
\usepackage[normalem]{ulem}

\definecolor{darkblue}{RGB}{0, 0, 139}
\definecolor{darkgreen}{RGB}{25, 125, 25}

\newcommand{\RR}[1]{\textcolor{black}{#1}}

\sethlcolor{blue!10}

\makeatletter
\newcommand{\daggerfootnotemark}{%
  \begingroup
  \protected@xdef\@thefnmark{$\dagger$}%
  \@footnotemark
  \endgroup
}

\newcommand{\daggerfootnotetext}[1]{%
  \begingroup
  \protected@xdef\@thefnmark{$\dagger$}%
  \@footnotetext{#1}%
  \endgroup
}
\makeatother

\begin{document}
 
\title{Performance of universal machine learning potentials in global optimization \RR{of inorganic crystal structures}}
\author{Edan T. Marcial}
\affiliation{Department of Physics, Applied Physics, and Astronomy, Binghamton University-SUNY, Binghamton, New York 13902, USA}
\author{Laxman Chaudhary}
\affiliation{Department of Physics, Applied Physics, and Astronomy, Binghamton University-SUNY, Binghamton, New York 13902, USA}
\author{Olesya Gorbunova}
\affiliation{Department of Physics, Applied Physics, and Astronomy, Binghamton University-SUNY, Binghamton, New York 13902, USA}
\author{Aleksey N. Kolmogorov}
\email{kolmogorov@binghamton.edu}
\affiliation{Department of Physics, Applied Physics, and Astronomy, Binghamton University-SUNY, Binghamton, New York 13902, USA}
\date{\today}  

\begin{abstract}
    Rapid development of universal machine learning potentials (uMLPs) and expansion of training data sets are reshaping the state of the art in atomistic simulation, highlighting the need for concurrent systematic benchmarking of their capabilities. Global optimization is among the most demanding uMLP applications because unconstrained exploration includes probing motifs not present in reference sets. We examined \RR{twelve pretrained} uMLPs in unconstrained evolutionary searches to assess whether these models can consistently predict complex \RR{nonmagnetic} crystal structure ground states \RR{under ambient pressure} across diverse inorganic systems. Our findings demonstrate that the considered M3GNet, MACE\RR{-MATPES, MACE-mh-1}, SevenNet, EquiformerV2\RR{, EquiformerV3}, MatterSim, GRACE, eSEN, Orb-v3, PET-MAD\RR{, and PET-OAM} models span a wide performance range, from near {\it ab initio} to essentially non-predictive, in their ability to resolve competing phases within low-energy basins. Additional tests on hcp-Zn, MB$_4$ (M = Cr, Mn, and Fe), and LiB$_{y}$ ($y\approx 0.9$) ground states reveal that several uMLPs capture fine energy differences arising from subtle electronic structure features.
\end{abstract}

\maketitle

\section{\label{sec:intro}Introduction}

Machine learning interatomic potentials (MLPs) have earned a place in mainstream computational materials research as efficient alternatives to electronic structure calculations. The first models capable of describing arbitrary atomic environments, the Behler–Parrinello neural networks (NNs) and the Gaussian approximation potentials (GAPs), demonstrated that density functional theory (DFT)-level accuracy can be retained in simulations where system sizes and times scales are extended by orders of magnitude~\cite{Behler2007,Bartok2010}. In the following years, MLPs were applied to a broad range of tasks, including molecular dynamics \RR{(MD)} simulations of phase stability, thermal transport, and mechanical response, geometry optimizations of clusters, surfaces, and extended defects, and transition state searches of diffusion and reaction pathways~\cite{ Deringer2019,Hart2021,Mishin2021}.

The impact of MLPs on crystal structure prediction was less immediate. Many early studies focused on elemental materials, {\it e.g.}, B, C, Si, and P, and either reproduced known ground states or identified metastable polymorphs~\cite{Huang2018,Podryabinkin2019,Behler2008,Deringer2018}. The first examples of new thermodynamically stable compounds found with MLP-accelerated global searches appeared only in the late 2010s~\cite{ak37,ak47,ak51,Ouyang2015,Gubaev2019}. The main challenge was the construction of practical potentials that could resolve small energy differences between competing structures while remaining reliable in previously unexplored regions of configuration space. This was addressed, in particular, by extending models from elements to multicomponent systems and by introducing automated workflows for iterative reference data generation and model parametrization~\cite{ak34,ak41,Dolgirev2016,Artrith2011}. The resulting MLPs offered genuine acceleration, leading to the identification of new complex M-Sn~\cite{ak47,ak51} and SiH$_4$~\cite{Pickard2022} ground states overlooked in prior searches at the DFT level. Despite these successes, the need to build robust system-specific models has limited the widespread use of MLPs in routine crystal structure prediction.

Recently introduced universal MLPs (uMLPs) represent a conceptually different approach. Trained on large and broadly representative materials databases, these models enable users to either bypass system-specific parametrization altogether or carry out further fine-tuning for target applications. uMLPs have already been used for large-scale screening of stability, transport properties, and finite-temperature behavior and have been benchmarked in several studies~\cite{Chen2022, Pota2024, Sharma2025, Yu2024, Loew2025, Riebesell2025, Chiang2025, Tahmasbi2025, Peng2026, Loew2025B, Lyngby2024}. In particular, Yu {\it et al.}~\cite{Yu2024} assessed four uMLPs on equation-of-state curves, relaxed geometries, and formation energies for extensive sets of crystalline materials. Riebesell {\it et al.}~\cite{Riebesell2025} developed the Matbench Discovery framework to evaluate property predictions, such as relaxed formation energies and distances to the convex hull, from unrelaxed structures. Chiang {\it et al.}~\cite{Chiang2025} presented the MLIP Arena system, which compares the performance of nine uMLPs across equation-of-state curves and \RR{MD} simulations, focusing on physically grounded performance independent of DFT references. Tahmasbi {\it et al.}~\cite{Tahmasbi2025} extended these evaluations to global structure searches by using a minima hopping algorithm to probe the potential energy surfaces \RR{(PES)} of unary materials. These studies establish that current uMLPs are sufficiently accurate for local optimization of pre-defined structure prototypes and near-equilibrium properties as well as applicable to global optimization of elemental systems, but their performance in unconstrained structure searches for stable chemically diverse inorganic compounds has yet to be explored systematically.

\begin{table*}[t]
\caption{Overview of uMLPs evaluated in this work, with additional information given in Section~\ref{mlp}. For each model we report the publication year, model architecture, featurizing basis functions, training data source, reference DFT method, training set size, and the parameter count. RBF, SBF, and GK stand for radial basis functions, spherical Bessel functions, and Gaussian kernels, respectively. Blank entries represent data not reported or readily available.}
\label{umlp}
\centering
\footnotesize 
\renewcommand{\arraystretch}{1.2}

\renewcommand{\tabularxcolumn}[1]{m{#1}} 
\newcolumntype{L}{>{\raggedright\arraybackslash}X}
\newcolumntype{C}{>{\centering\arraybackslash}X}
\newcolumntype{R}{>{\raggedleft\arraybackslash}X}

\begin{tabularx}{\textwidth}{@{} 
    >{\hsize=0.98\hsize}L 
    >{\hsize=0.4\hsize}C 
    >{\hsize=0.5\hsize}C 
    >{\hsize=0.9\hsize}L 
    >{\hsize=1.2\hsize}L 
    >{\hsize=1.5\hsize}L 
    >{\hsize=0.4\hsize}L 
    >{\hsize=0.63\hsize}R 
    >{\hsize=0.5\hsize}R 
@{}}
\toprule
uMLP & ID\phantom{$^{*}$} & Year & uMLP & Featurizing & Dataset & DFT & Structures & Parameters \\
version & adopted\phantom{$^{*}$} & reported & architecture & basis functions & sources & functional & $(\times 10^6)$ & $(\times 10^6)$ \\
\midrule
M3GNet       & MG\phantom{$^{*}$}& 2022 & GNN          & Bessel RBF $Y_{\ell}^{m}$ & MP                & PBE    & 0.2\phantom{0}   & 1.1\phantom{00} \\
MACE\RR{-MATPES}         & \RR{MC$^*$}  & 2023 & Eqv. MPNN    & Bessel RBF $Y_{\ell}^{m}$ & MATPES-PBE        & PBE    &                  & 19.8\phantom{00} \\
\RR{MACE-mh-1}         & \RR{MC\phantom{$^{*}$}} & \RR{2025} & \RR{E(3) GNN} & \RR{Bessel RBF $Y_{\ell}^{m}$} &  \RR{OMat24} &    \RR{PBE/PBE+U}    & \RR{110}\phantom{.0}   &      \RR{19.8} \phantom{.0}       \\
SevenNet-MF  & SN\phantom{$^{*}$} & 2024 & E(3) NequIP  & Bessel RBF $Y_{\ell}^{m}$ & OMat24 MPtrj sAlex & PBE    & 113\phantom{.0}  & 25.7\phantom{00} \\
EquiformerV2 & \RR{EQ{$^*$}}  & 2024 & E(3)/SE(3)   & \RR{Gaussian} RBF $Y_{\ell}^{m}$ & OMat24 MPtrj sAlex & PBE    & 102\phantom{.0}  & 98.4\phantom{00} \\
\RR{EquiformerV3}         & \RR{EQ\phantom{$^{*}$}} & \RR{2026} & \RR{SE(3)} & \RR{Gaussian RBF $Y_{\ell}^{m}$} &   \RR{OMat24 MPtrj sAlex}    &   \RR{PBE}  & \RR{113\phantom{.0}}            & \RR{30\phantom{00}}\\
MatterSim    & MS\phantom{$^{*}$} & 2024 & M3GNet Graph & SBF $Y_{\ell}^{m}$ GK & MatterSim         & PBE    & 17\phantom{.0}   & 4.6\phantom{00} \\
GRACE-2L     & GR\phantom{$^{*}$} & 2025 & ACE Graph    & Chebyshev RBF $Y_{\ell}^{m}$ & OMat24          & PBE    &                  & 25.2\phantom{00} \\
eSEN     & EN\phantom{$^{*}$} & 2025 & E(3)/SE(3)   & \RR{Gaussian} RBF \RR{$Y_{\ell}^{m}$}       & OMat24            & PBE    & 10\phantom{.0}   & 30.2\phantom{00} \\
Orb-v3       & OR\phantom{$^{*}$} & 2025 & GNS          & Bessel RBF $Y_{\ell}^{m}$ & OMat24            & PBE    & 55\phantom{.0}   & 25.5\phantom{00} \\
PET-MAD      & \RR{PT{$^*$}}  & 2025 & GNN          & Machine learned           & MAD               & PBEsol & 0.1\phantom{0}   & 3.5\phantom{00} \\
\RR{PET-OAM} & \RR{PT\phantom{$^{*}$}} & \RR{2026} & \RR{GNN} & \RR{Machine learned} &  \RR{OMat24 MPtrj sAlex MAD}     &  \RR{PBE}   &    \RR{113\phantom{.0}}              & \RR{730}\phantom{00}\\
\bottomrule
\end{tabularx}
\end{table*}

In this work, we assess the performance of the latest generation of uMLPs \RR{as well as three legacy models} in global structure searches in multicomponent systems. The workflow complements existing benchmarks by probing the surrogate \RR{PES} beyond the configurations represented in the training datasets, as we investigated materials with both well-established~\cite{Mair1999,ak21,Becher1962, Hlukhyy2004, Eguchi2011,Howard1991,Henn1996,Ahn2016,Berthold1976,Becker2000} and recently proposed~\cite{ak47,ak51,ak54} ambient-pressure ground states. The
\RR{twelve} selected models, M3GNet~\cite{Chen2022}, MACE-\RR{MATPES}~\cite{Batatia2022,Batatia2025}, \RR{MACE-mh-1}\RR{~\cite{Batatia2025b}}, SevenNet~\cite{Park2024,Kim2024},  EquiformerV2~\cite{Liao2024,Barroso-Luque2024, Liao2023,Chanussot2021}, \RR{ EquiformerV3~\cite{Liao2026}, }MatterSim~\cite{Yang2024}, GRACE~\cite{Lysogorskiy2024, Lysogorskiy2025}, eSEN~\cite{Barroso-Luque2024,Fu2025}, Orb-v3~\cite{Rhodes2025,Neumann2024}, PET-MAD~\cite{Mazitov2025}, 
\RR{and PET-OAM~\cite{Bigi2026}} span various architectures and vastly different training sets, but nearly all of them were fitted to the same DFT approximation (see Table~\ref{umlp}). To determine their suitability for ground state searches across materials with different chemistries and bonding types, we used the uMLPs as out-of-the-box surrogate models combined with an evolutionary algorithm~\cite{ak41}. Within this framework, which previously enabled the prediction of new thermodynamically stable compounds~\cite{ak47,ak51}, the small pools of identified low-energy candidate structures are re-examined with the reference DFT approximation. We also consider three challenging cases, hcp-Zn with an anomalous $c/a$ ratio, distorted MB$_4$ derivatives (M = Cr, Mn, and Fe), and off-stoichiometric LiB$_y$ phases, in which electronic features are known to play a key role in defining the ground state structure. \RR{By simulating all selected systems with three DFT flavors}, the Perdew–Burke–Ernzerhof (PBE)~\cite{Perdew1996}, PBE for solids (PBEsol)~\cite{PBEsol}, and regularized-restored strongly constrained and appropriately normed (r2SCAN)~\cite{Furness2020} exchange-correlation functionals, \RR{we contextualize the uMLP errors and} demonstrate that some models, most notably \RR{EquiformerV3, PET-OAM, and} eSEN, reproduce the reference method with accuracy better than the systematic spread among these DFT approximations.

\section{Methodology}
\subsection{DFT Calculations}
\label{dft}

All DFT calculations were performed with the Vienna {\it ab initio} simulation package (VASP)~\cite{Kresse1993,Kresse1994,Kresse1996A,Kresse1996B} and projector-augmented wave potentials~\cite{Blochl1994A} with the maximum number of semi-core electrons for each element. We used a 500 eV plane-wave cutoff and dense Monkhorst–Pack~\cite{Monkhorst1976} k-meshes with $\Delta k \leq 2\pi \times 0.025$ \AA$^{-1}$ to ensure good numerical convergence. All final energies were evaluated with the tetrahedron integration method with Bl\"{o}chl corrections~\cite{Blochl1994B} for fully optimized unit cells. As the default DFT method \RR{for nearly all uMLPs (see Table~\ref{umlp})}, we employed the PBE exchange-correlation functional~\cite{Perdew1996} within the generalized gradient approximation (GGA)~\cite{Langreth1983}. 
For examining systematic DFT errors, we also carried out calculations with the PBEsol~\cite{PBEsol} functional, the r$^2$SCAN functional~\cite{Furness2020} that is currently adopted in the Materials Project~\cite{Jain2013} for its improved description of materials formation energies, and four different parametrizations~\cite{Klime2009,Klime2011,Ning2022,Chakraborty2020} that treat van der Waals (vdW) interactions important in layered materials.

\subsection{uMLPs}
\label{mlp}

Pretrained models were taken directly from publicly available repositories and were not fine-tuned in this study in order to evaluate their baseline performance. M3GNet (denoted as MG henceforth) is a pioneering model trained on three-body interactions across 89 elements~\cite{Chen2022}. \RR{MACE-MATPES} (\RR{MC$^*$}) utilizes the atomic cluster expansion as a descriptor and performs efficient calculations due to its streamlined message-passing architecture~\cite{Batatia2022,Batatia2025}. \RR{MACE-mh-1 (MC) is based on MACE-OMAT-0, which shares an architecture with MACE-MATPES but differs in training data, and adds higher order messages, nonlinear blocks, and a multi-head replay post-training strategy~\cite{Batatia2025b}}. SevenNet (SN) maintains equivariance, incorporates the NequIP \RR{framework}~\cite{Batzner2022}, and uses a combination of low- and high-fidelity data in its training~\cite{Park2024,Kim2024}. EquiformerV2 (\RR{EQ{$^*$}}) is an equivariant model that \RR{incorporates} eSCN convolutions~\cite{Passaro2023} for higher degree representations of atomic environments~\cite{Liao2024,Barroso-Luque2024,Liao2023,Chanussot2021}. \RR{EquiformerV3 (EQ) further develops this architecture by introducing SwiGLU-\textit{S}$^{2}$ activations for strict equivariance on coarser grids and a gradual radius cutoff that helps produce smoothly varying PES~\cite{Liao2026}.} MatterSim (MS, v1.0.0--5M) uses active learning, uncertainty-aware data sampling, and a training dataset containing off-equilibrium structures under a variety of temperatures and pressures~\cite{Yang2024}. GRACE (GR) uses equivariant message passing and extends the atomic cluster expansion to recursively evaluated star- and tree-like basis functions, constructing a complete basis for describing atomic interactions~\cite{Lysogorskiy2024,Lysogorskiy2025}. eSEN (EN) is a message-passing NN with an Equiformer-like architecture~\cite{Liao2024} designed to strictly maintain equivariance and a smoothly varying \RR{PES}~\cite{Barroso-Luque2024,Fu2025}. The conservative\RR{, unlimited neighbor} Orb-v3 (OR) variant is a roto-equivariant graph network simulator (GNS) that builds off of the previous generation of direct-force Orb models~\cite{Rhodes2025,Neumann2024}. PET-MAD (\RR{PT{$^*$}, v1.0.2}) is a rotationally unconstrained, transformer-based model trained on the Massive Atomic Diversity (MAD) dataset~\cite{Mazitov2025}. \RR{The recently introduced PET-OAM (PT) updates the PET \RR{framework} by passing node features up GNN layers, introducing a new direct force head, and increasing the number of node features, allowing for dramatically more parameters (see Table~\ref{umlp}) at minimal cost~\cite{Bigi2026}}. Among these uMLPs, EQ$^*$  is the only explicitly non-conservative model that calculates forces directly rather than as an energy gradient. 

In the present tests, local structure relaxations were performed with MatGL~\cite{Ko2025} in MG simulations or with the Atomic Simulation Environment (ASE) for all other uMLPs~\cite{Hjorth2017}, and the accuracy of each uMLP's predictions was evaluated against the DFT functional used in its training, {\it i.e.}, PBEsol for PET-MAD and PBE for the rest (see Table~\ref{umlp}). Importantly, none of the tested properties involved absolute DFT energies, which made results independent of the atomic reference values in pseudopotentials, particularly relevant in the case of the PT$^*$ baseline method~\cite{Mazitov2025}. \RR{For clarity, the main-text comparisons focus primarily on the nine latest uMLPs, whereas the detailed performance of the legacy variants, PT$^*$, MC$^*$, and EQ$^*$, is reported in the Supplemental Material.}

Information on uMLP relative efficiencies \RR{in GPU-based high-throughput and MD workflows} can be found in previous studies, e.g., Refs.~\cite{Mazitov2025,Peng2026}. \RR{All present uMLP global searches, requiring sequential generation and local optimization of structures, were carried out on CPU nodes. Our single-core CPU benchmark for fixed-length relaxations of crystal structures with 10--16 atoms per unit cell, shown in Fig. S1, is qualitatively consistent with the reported GPU trends: runtimes depend strongly on model architecture and chemical system, spanning a factor of over 30 between MG and EQ$^*$ in our tests.}

\subsection{Global search benchmarking protocol}
    \label{testbed}
    
Efficient acceleration of {\it ab initio} global structure optimization depends on surrogate models meeting two key requirements~\cite{ak40,ak41,ak47,Tahmasbi2025}. First, a uMLP must be robust enough that the PES is not dominated by spuriously overstabilized unphysical structures and that the true DFT ground state can be reached from a uMLP minimum structure via a barrierless trajectory. Second, it must be sufficiently accurate that the uMLP-based energy ranking retains, within a tractably small pool, at least one candidate that converges to the true ground state upon DFT re-optimization. Systematic comparison of the uMLPs’ performance is not straightforward due to the inherently stochastic nature of global optimization methods. The suitability of uMLPs for finding ground states with no structural input was benchmarked with the following protocol.
    
The starting point involved carrying out typical zero-temperature fixed-composition evolutionary searches appropriate for locating small- to medium-sized ground states of crystalline solids~\cite{ak47,ak51}. Populations of 100 structures with a given number of atoms were seeded randomly and evolved over 100 generations with standard operations implemented in the module for {\it ab initio} structure evolution ({\small MAISE})~\cite{ak41}\RR{:} mutation of individual members (20\%), crossover of two parent structures (60\%), and injection of additional random seeds (20\%). \RR{Such searches are usually sufficient but not foolproof for locating ground states in this size range. When the reference global minimum was missed, we carried out up to four additional attempts and report in Table S1 the generation at which that structure entered the population. The 10$^4$} uMLP-level local optimizations \RR{in each run} proceeded until the maximum atomic force fell below 0.05~eV/\AA, which \RR{enabled reliable sampling of different basins by detecting and eliminating duplicate motifs} based on the similarity in energy and structural fingerprint~\cite{ak16,ak41}. To remove any underconverged duplicates, all visited minima up to 20 meV/atom above the best candidate and with the similarity fingerprint scalar product below 0.92, controlled by the SCUT flag in {\small MAISE}, were optimized with the uMLP with a finer 0.001~eV/\AA\ tolerance and filtered again with the 0.92 threshold. For compounds where few minima were found after filtering, we increased the energy window by 10 meV/atom and repeated the filtering step until more than 10 minima were found or the energy window reached 100 meV/atom ({\it e.g.}, for LiB$_3$). Inspection of the pools provided information on the quality of the PES fit for individual uMLPs, judged by the presence of minima significantly below the global one and an overabundance of \RR{near-degenerate lattice decorations or} low-symmetry satellite minima around structures improperly modeled as dynamically unstable. \RR{Accordingly}, we used a more restrictive \RR{0.80 (Li$_3$Sn) or 0.85 (Na$_2$CN$_2$)} SCUT flag for our selection criterion. 
    
The variance in the number of candidate structures per uMLP pool complicated consistent quantitative assessment of the models’ ability to select competitive candidates in relevant regions of the configuration space. To account for the non-deterministic sampling of low-energy basins and ensure that each uMLP test pool included the ground state \RR{for each compound}, we merged \RR{selected minima from one evolutionary run for each uMLP (as detailed in Table S2) so that cases requiring additional tries did not contribute disproportionately to the aggregate. When multiple attempts were available, we retained the run reaching the lowest-energy uMLP minimum.} Subsequent uMLP-level relaxation (0.001~eV/\AA\ tolerance) and filtering (0.92 SCUT) of the \RR{merged sets} generated comparable pools for each classical model. \RR{In addition to the SCUT filter criteria, we also only collected minima up to 100 meV/atom above the lowest energy candidate to avoid the appearance of high energy artifacts that could skew our evaluation. This energy window filter led to an average of 17\% of structures in uMLP relaxed pools being filtered out, leaving only relevant low energy candidates for further analysis.}
\RR{Finally, candidates in each uMLP pool were re-optimized with the reference DFT method, generally after conservative 0.01-tolerance MAISE symmetrization to remove numerical noise.}
    
The fingerprint scalar product and the DFT energy difference between the uMLP- and DFT-relaxed configurations, defining structure and energy proximity metrics, respectively, indicated how close the competing minima were on the surrogate and reference PESs. The ranking accuracy was gauged using an RMSE defined as
    \begin{equation}
    \text{Ranking RMSE} = \left[ \frac{\sum_{n=1}^{N} \left( E^{\text{uMLP}}_n - E^{\text{DFT}}_n - \Delta \overline{E} \right)^2}{N-1} \right]^{\frac{1}{2}}.
    \nonumber
    \end{equation}
Here, $N$ is the number of structures in each pool, $\Delta \overline{E} = \overline{E}^{\text{uMLP}} - \overline{E}^{\text{DFT}}$ is the difference between pool averages, and the $N-1$ factor accounts for the Bessel correction. This metric differs from the typically reported overall model error in that it eliminates the average shifts irrelevant to relative stability, evaluates energies of configurations fully relaxed within their respective methods rather than identical snapshots, and probes accuracy only in the low-energy basins important in global structure searches. 

\RR{Metrics for ground state screening efficiency are important but not straightforward to define, in part because establishing whether multiple uMLP minima converge to the same DFT ground state is sensitive to the tolerance limits of the reference calculations. We therefore constructed the ground state hit set for each compound as all shortlisted minima of a given uMLP that, after DFT re-optimization, were assigned to this endpoint using the structural fingerprint cutoff. The uMLP candidates were ordered by their uMLP energies, and $\mathrm{rank}_i$ was defined as the earliest position in this ordered list among members of the hit set for compound $i$. The mean reciprocal rank was then calculated as
\[
\mathrm{MRR}=\frac{1}{N_c}\sum_{i=1}^{N_c}\frac{1}{\mathrm{rank}_i},
\]
where $N_c$ is the number of considered compounds. The uncertainty in the average MRR was estimated from 95\% Student $t$ confidence intervals over the considered compounds.}

\RR{The size of the hit set defined the ground state multiplicity (GSM). Large GSM values indicate that the shortlist contains many ground state satellites, and numerous DFT re-optimizations are spent rediscovering the known minimum \RR{instead of screening distinct metastable candidates}. Reliable ordering of metastable candidates would be valuable for screening workflows~\cite{Fasoulis2024,Qu2024}, especially when low lying phases beyond the 0~K ground state are relevant~\cite{ak47}. We considered $K$ ranking metrics, including the normalized discounted reciprocal rank (nDRR$@K$), but did not include them for two reasons. First, most successful uMLPs already recovered the ground state at rank one, so finite $K$ scores would add little beyond MRR for those models. Second, automated hit list construction beyond the known ground state would suffer from accumulated ambiguities, as fingerprint-based assignments of shallow DFT minima are susceptible to numerical noise and cannot be readily checked in regions of genuinely fragmented reference PES landscapes.}

\RR{Operationally, ranking RMSE, MRR, and GSM provided complementary information on the relative stability within the candidate shortlists, rank efficiency in finding the reference ground state for each compound, and DFT screening overhead from redundant global minimum hits.}
    
\section{Evolutionary optimization}
\label{evos}

\begingroup
\begin{table*}[!ht]
\centering
\footnotesize
\caption{uMLP energy difference of best candidates from evolutionary searches relative to reference ground states given in meV/atom.  \RR{The formula and Pearson symbol for each reference are provided in the first two rows. In the case of Na$_2$CN$_2$, the mS10 experimental structure favored by PBEsol is used as the reference for PT$^{*}$ while the tI10 PBE artifact is used as the reference for all other models.} The relative rank of reference structures found in each pool is given as a positive superscript \RR{(degenerate structures with energy differences $< 0.1$ meV/atom are assigned the same rank).} Searches where the \RR{reference} global minimum was not found have a positive energy difference or a zero superscript.}
\label{main}

\newlength{\evosfirstcol}
\newlength{\evosdatacol}
\settowidth{\evosfirstcol}{MC$^{{*}}$}
\addtolength{\evosfirstcol}{2pt}
\setlength{\evosdatacol}{\dimexpr(\textwidth-\evosfirstcol)/11\relax}

\newcolumntype{V}{>{\centering\arraybackslash}p{\evosfirstcol}}
\newcolumntype{E}{S[
  table-format=-3.1,
  table-column-width=\evosdatacol,
  table-number-alignment=center,
  table-text-alignment=center,
  table-space-text-post={$^{27}$}
]}

\setlength\tabcolsep{0pt}
\begin{tabular}{@{}V*{11}{E}@{}}
\hline\hline \noalign{\vskip 1mm}
{uMLP} & {\phantom{00}LiB$_3$} & {\phantom{0}Be$_4$B} & {\phantom{00}TiO$_2$} & {\phantom{00}MgIr$_2$} & {\phantom{00}Pd$_5$Sn$_3$} & {\phantom{00}Li$_3$Sn} & {\phantom{00}Na$_2$CN$_2$} & {\phantom{0}Na$_2$IrO$_3$} & {\phantom{0}Si$_3$CaPt} & {\phantom{0}MgB$_3$C$_3$} & {\phantom{0}AgClO$_4$}\\
{ID\phantom{{$^{*}$}}}    & {\phantom{00}tP16}    & {\phantom{0}tP10}    & {\phantom{00}mS24}     & {\phantom{00}hP12}     & {\phantom{00}mS32}     & {\phantom{00}hR48}     & {\RR{\phantom{0}tI10 (mS10)}} & {\phantom{0}mS24} & {\phantom{0}tI10} & {\phantom{0}oI28} & {\phantom{0}tI12}\\
\noalign{\vskip 2pt}
\hline
\noalign{\vskip 1pt}
EN\phantom{{$^{*}$}} & 0.0 & 0.0 & 0.0 & 0.0 & 0.0 & 0.0 & 0.0 & 0.0 & 0.0 & 0.0 & -187.4{$^{0}$}\\
EQ{$^{*}$} & 0.0 & 0.0 & 0.0 & 0.0 & 0.0 & 0.0 & -1.0{$^0$} & 0.0 & 0.0 & 15.7 & -185.3{$^{0}$} \\
EQ\phantom{{$^{*}$}} & 0.0 & 0.0 & 0.0 & 0.0 & 0.0 & 0.0 & 0.0 & 0.0 & 0.0 & 28.4 & -180.0{$^{0}$} \\
OR\phantom{{$^{*}$}} & 0.0 & 0.0 & 0.0 & 0.0 & 0.0 & 0.0 & 0.0 & 0.0 & 0.0 & 0.0 & -189.9{$^{0}$}\\
SN\phantom{{$^{*}$}} & 0.0 & 0.0 & 0.0 & 0.0 & 0.0 & 0.0 & 0.0 & 0.0 & 0.0 & 0.0 & -205.7{$^{0}$}\\
GR\phantom{{$^{*}$}} & 0.0 & 0.0 & -3.4{$^{2}$} & 0.0 & 0.0 & -2.6{$^{4}$} & -0.4{$^3$} & 0.0 & 0.0 & 0.0 & -210.9{$^{0}$}\\
MC{$^{{*}}$} & 0.0 & 0.0 & -37.0{$^{27}$} & 0.0 & 0.0 & -7.2{$^{2}$} & 0.0 & 0.0 & 0.0 & -45.1{$^{0}$} & {-1e4$^{0}$}\\
MC\phantom{{$^{*}$}} & 0.0 & 0.0 & 0.0 & 0.0 & 0.0 & 0.0 & -4.3{$^2$} & 0.0 & 0.0 & 0.0 & -205.6{$^{0}$}\\
MS\phantom{{$^{*}$}} & 0.0 & 0.0 & 0.0 & 0.0 & 0.0 & 0.0 & 0.0 & 0.0 & 0.0 & 0.0 & -179.0{$^{0}$}\\
PT$^{*}$ & 0.0 & 0.0 & -5.2{$^{2}$} & 0.0 & 0.0 & 0.0 & 0.0 & 0.0 & 0.0 & 0.0 & -157.0{$^{0}$}\\
PT\phantom{{$^{*}$}} & 0.0 & 0.0 & 0.0 & 0.0 & 0.0 & 0.0 & 0.0 & 0.0 & 0.0 & 25.3 & -160.3{$^{0}$}\\
MG\phantom{{$^{*}$}} & -47.6{$^{0}$} & -14.5{$^{0}$} & -9.4{$^{0}$} & -29.7{$^{0}$} & -3.8{$^{20}$} & -14.9{$^{0}$} & -56.4{$^0$} & 0.0 & -14.8{$^{0}$} & -180.3{$^{0}$} & -220.8{$^{0}$}\\
\hline\hline
\end{tabular}
\end{table*}
\endgroup
    
Evolutionary searches were performed for 12 inorganic nonmagnetic compounds with bonding environments ranging from extended metal-stabilized covalent-ionic frameworks to interpenetrating metal-metalloid sublattices and structural complexities ranging from 5 to 16 atoms per primitive unit cell \RR{(see Table S1)}. The inclusion of Li$_3$Sn, Pd$_5$Sn$_3$, and MgB$_3$C$_3$ with recently proposed prototypes~\cite{ak47,ak51,ak54} served to probe whether the uMLPs can recognize unfamiliar, non-sampled motifs as ground states.

Table~\ref{main} summarizes the qualitative outcome of the searches \RR{(see Supplementary Listing 1 for MAISE settings)}. For each model and compound, we report the uMLP energy of the corresponding lowest-energy candidate relative to the ground state of the reference DFT method and the position of the DFT ground-state structure within the uMLP-ranked pool as a superscript. 
\RR{Negative entries identify competing minima favored by the uMLP over the reference phase, whereas small positive superscripts mean that this phase would be recovered from a shallow post-search DFT shortlist.}
\RR{In the particularly challenging MgB$_3$C$_3$ case, five runs were insufficient for some uMLPs despite the TETRIS-based generation~\cite{ak56} of initial structures containing B-C building blocks \RR{(see Supplementary Listing 2)} used to accelerate the convergence. We would like to emphasize that this} dataset is not statistically sufficient to disentangle whether missed ground states reflect surrogate deficiencies or seed/trajectory dependence in the evolutionary exploration. \RR{However, the investigation did identify two clear model- and compound-level outliers, MG and AgClO$_4$, as discussed below.}

Evolutionary searches with nearly all uMLPs reached the global minima for the majority of considered compounds, mostly independent of the primitive unit cell size or perceived complexity. For example, the relatively large LiB$_3$ ground state (tP16) is comprised of compact B$_6$ units that recur frequently across metal boride chemistries~\cite{ak28}, which is apparently well reproduced as a broad basin of attraction on surrogate PESs. Likewise, the low-symmetry Na$_2$IrO$_3$ (mS24) features a typical layered arrangement of edge-sharing IrO$_6$ octahedra, which tends to constrain the accessible low-energy motifs. The recently proposed Li$_3$Sn (hR48) and Pd$_5$Sn$_3$ (mS32) ground states with new structure types turned out not to pose particular difficulty, as their local environments can be viewed as ordered decorations of familiar metallic and metal-metalloid coordination motifs~\cite{ak47,ak51}, and the uMLPs learned the underlying stability trend well enough to identify these phases as global minima. \RR{Interestingly, about half of the uMLPs favored a distorted mS32-Pd$_5$Sn$_3$ variant, indicating that reducing the reported $C2/c$ symmetry to $Cc$ lowers the energy by 0.1--0.3 meV/atom. PBE and PBEsol yield similar energy gains along a $\Gamma$-point eigenvector in the conventional unit cell, associated with a single imaginary mode that breaks inversion symmetry through buckling of the Pd-Sn framework. However, the extremely small energy change and minor atomic displacements of about 0.1~\AA\ do not necessarily imply that the distortion would be detectable, as it could be suppressed by anharmonic effects. Hence, despite the similarity factor of 0.74 being below our typical threshold, we treated the two phases as identical in our MRR and GSM metric calculations.}

The few misses, either failing to access the global-minimum basin or misranking closely competing minima, were concentrated in systems that are challenging even at the DFT level. TiO$_2$ exhibits a rich polymorphism, with the relative stability of low-energy phases being highly sensitive to the treatment of dispersive and correlation effects~\cite{Zhang2019}. This complexity of the PES helps frame why two uMLPs reversed the stability order of anatase (tI12) and the PBE-level ground state bronze (mS24) and why another did not allow the searches to log the ground state as a viable candidate. The uMLP-based exploration successfully recovered other known phases, namely, rutile (tP6), columbite (oP12), and baddeleyite (mP12), which indicates that the models correctly distinguished the PBE ground state from a realistic line-up of candidates. 

Global optimization of AgClO$_4$ with all 12 uMLPs produced numerous spuriously deep minima containing O$_2$ molecules, such as mS24 given in the Supplemental Material \RR{(Fig. S2)}, starkly contrasting the accepted tI12 ground state built from ClO$_4$ tetrahedra. This pathology indicates that the considered uMLPs are not sufficiently familiar with molecular oxygen to disfavor its appearance in solid-state phases, which should be straightforward to fix by retraining the models on datasets that include this motif. \RR{Since the failure occurs across diverse architectures, AgClO$_4$ proves to be an out-of-distribution stress test of the released models rather than a conventional low-energy ranking error. In an uncertainty-aware or active-learning workflow, such chemically implausible configurations would be natural triggers for early DFT evaluation and targeted retraining before they dominate evolutionary sampling.} Similar measures may be needed when exploring related systems to prevent the unphysical phase separation into notably stable low-dimensional units, such as nitrogen or hydrogen molecules.

\begin{figure}[t!]
    \centering
    \includegraphics[width=0.45\textwidth]{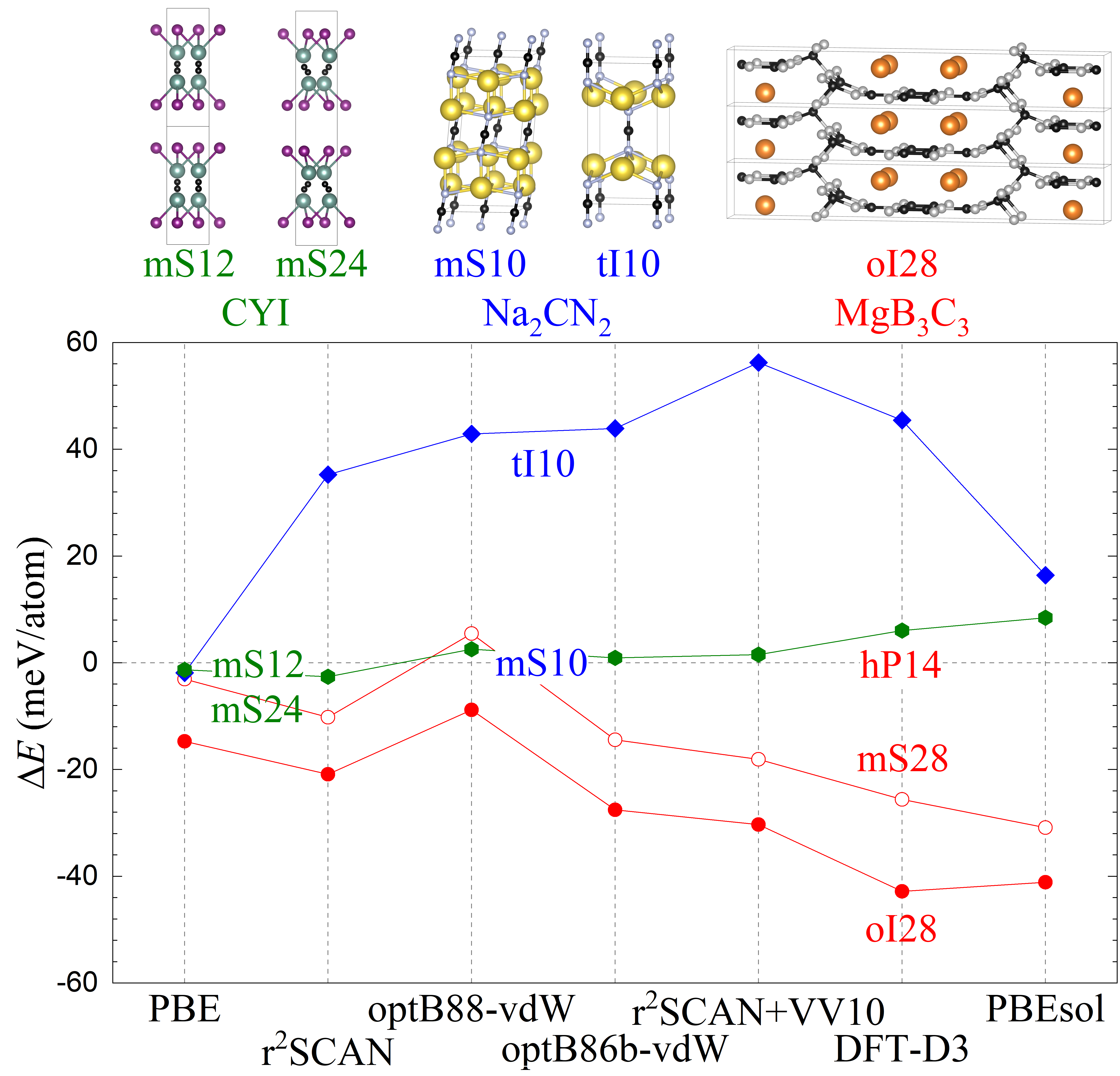}
        \caption{ Stability of the \RR{mS24-CYI}, tI10-Na$_2$CN$_2$, and oI28-MgB$_3$C$_3$ phases identified in this work relative to the reported \RR{mS12-CYI}, mS10-Na$_2$CN$_2$, mS28-MgB$_3$C$_3$, and hP14-MgB$_3$C$_3$ evaluated with common DFT functionals.}
        \label{fin}
\end{figure}

\RR{In addition to} the primary intended benchmarking objective, our global searches \RR{with several uMLPs unexpectedly identified lower-energy candidates at the reference level: two apparent semi-local functional artifacts and a robustly better polymorph.} For Na$_2$CN$_2$, a substantially less dense packing of CN$_2$ units in a tI10 candidate (17.8~\AA$^3$/atom) than in the known mS10 structure (13.4~\AA$^3$/atom) improved the PBE-level stability by 1.9 meV/atom (see Fig.~\ref{fin}). However, all other tested functionals soundly disfavor tI10 relative to mS10 by 16--56 meV/atom, which makes the outlier likely PBE-specific. \RR{For CYI, the experimentally reported mS12 phase~\cite{Henn1996,Ahn2016} is reproduced by PBEsol and the vdW-corrected functionals, while PBE and r$^2$SCAN favor a related mS24 variant with tilted C$_2$ dimers (see Fig.~\ref{fin}). This near degeneracy is expected for a layered compound in which shifted Y$_2$I$_2$ sheets and C$_2$ dimers are sensitive to dispersive interactions. In fact, this is reflected in the different stacking orders found among the collected uMLP minima and a consistent} 1--2~\AA\ overestimation of the interlayer spacing, an artifact inherited from the underlying semi-local DFT approximations that misestimate the measured values by 0.2--0.5~\AA. \RR{Due to the abundance of distorted, nearly degenerate structures that made the minima ranking unreliable, we excluded this compound from Table~\ref{main} and used it only in the calculation of ranking RMSE and proximity metrics.} For MgB$_3$C$_3$, proposed to be a high-$T_{\textrm c}$ superconductor in hP7~\cite{Pham2023} or hP14~\cite{ak54} honeycomb layer structures potentially accessible through deintercalation of the known MgB$_2$C$_2$ compound, prior {\it ab initio} searches located an mS28 phase with an alternative layered morphology considered more stable by a set of DFT functionals~\cite{ak54}. The current uMLP-accelerated global optimization generated an oI28 polymorph, shown in Fig.~\ref{fin}, with a BC framework connected in 3D via sp$^3$ links uniformly favored by all tested DFT approximations.

The quantitative correspondence between surrogate and DFT low-energy landscapes was assessed \RR{using merged pools and} the three metrics described in Section~\ref{testbed}. Table S1 shows a relatively large dispersion in the number of candidate phases in each uMLP’s \RR{original set}, {\it e.g.}, between \RR{14 for MC$^*$ and 45 for EN in Li$_3$Sn,} which illustrates why aggregating relevant minima was necessary for evaluating the uMLP performance on a more comparable basis. \RR{MG was treated separately because its native pools were unreasonably large (see Table~S1).} To enable a controlled comparison of alternative DFT flavors and the MG model, we chose the representative \RR{merged} EN set. \RR{For MG, this choice is favorable because it excludes numerous artifacts generated in its own searches, but the metrics below show that its ranking fidelity remains poor.} The resulting test sets for 11 compounds (not including AgClO$_4$) \RR{were mostly comparable in size, varying between 500 and 870 structures, with the only notable exception (EQ$^*$) exceeding 2,000 candidates. The inflated EQ$^*$ pools contained a large number of distorted structure variants, which complicated duplicate removal and suggests practical inefficiency in evolutionary searches that rely on compact candidate lists. The similar trends obtained from simple and size-weighted averages, with weights $N{^\textrm{minima}_\textrm{compound}}/N{^\textrm{minima}_\textrm{total}}$, indicate that the conclusions were not controlled by the largest sets.}

\begin{figure}[t!]
       \centering
    \includegraphics[width=0.48\textwidth]{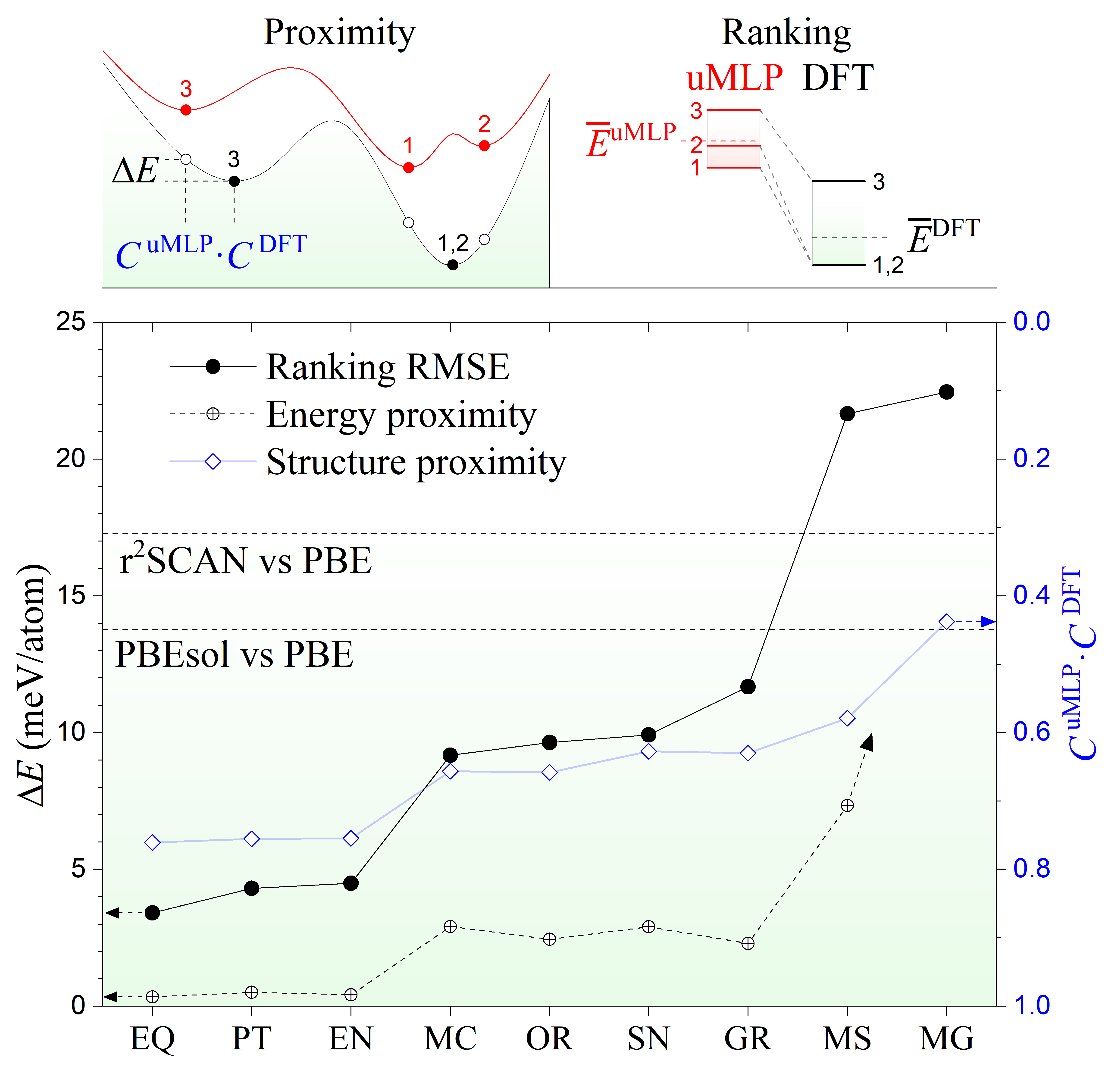}
        \caption{ Performance metrics of nine uMLPs on merged pools assessed relative to the reference DFT method, and averaged over 11 compounds, excluding AgClO$_4$. \RR{The comparison of EQ$^*$, PT$^*$, and MC$^*$ against their latest counterparts is presented in Fig. S3.} The schematics at the top clarify the definitions of the proximity \RR{metrics} and ranking \RR{RMSE} introduced in the text. The representative merged EN pools were used to evaluate the metrics for the alternative DFT approximations and MG, with dashed horizontal lines marking the ranking RMSE for PBEsol and r$^2$SCAN relative to PBE. The energy proximity average for MG exceeded 60 meV/atom (see \RR{Fig. S4} for further discussion).}
        \label{metrics}
\end{figure}

Fig.~\ref{metrics} summarizes the pool size-weighted metrics measuring the proximity of uMLP minima to the corresponding DFT counterparts and ranking fidelity within a low-energy shortlist. The changes in the \RR{radial distribution function} (RDF) fingerprint and the DFT energy upon re-optimization span 0.43--0.76 and 1--8~meV/atom, respectively, with \RR{EQ, PT, and} EN showing particularly small deviations. The fact that the average DFT relaxation produces only a modest and fairly systematic energy drop is important for downstream selection strategies that aim to improve effective ranking. For instance, our study on Au clusters~\cite{ak40} showed that a hybrid workflow, in which structures are relaxed with a NN potential while final energies are evaluated with single-shot DFT, can markedly improve convergence to the ground state, and that the added DFT cost during the evolutionary run can be offset by the reduction in pool sizes.

Some of the ranking RMSEs, ranging between 3 and 22~meV/atom, are notably lower than the overall errors typically reported for uMLPs, {\it e.g.}, around 20~meV/atom in MatBench tests~\cite{Riebesell2025}. This observation is consistent with the different functional purpose of this metric that involves only competing structures near the ground states and explicitly removes the average \RR{uMLP-DFT energy offset in each pool. Fig. S4} breaks down the ranking RMSE by compound and shows that the values are reasonably consistent across distinct chemical classes. The violin plots presented in \RR{Fig. S5} underscore a general absence of outliers that could dominate and skew the computed \RR{metric. Fig. S6 shows that the ranking RMSEs vary little between native and merged pools, differing within 1 meV/atom for most models.} The results show that relative energy accuracy correlates with the uMLP fitted-parameter count listed in Table~\ref{umlp}. \RR{Three models outperforming this trend are the medium-tier EQ and EN models with accuracies close to the large-sized EQ{$^*$} and PT models while the small-tier PT{$^*$}} approaches the level of the medium-sized uMLPs.

\RR{The computed MRR values of 0.82--0.91 for all but three uMLPs indicate remarkable consistency among most models in ranking the DFT ground state near the top of the candidate pool. The contribution from the artifact-dominated AgClO$_4$ case was set to zero. The confidence-interval half-widths of 0.18--0.25 do not allow for a precise ordering within this band. MC$^\ast$ and GR, with lower values of $0.70\pm0.25$ and $0.71\pm0.25$ respectively, suggest less consistent ranking fidelity, although their intervals still overlap with the main group. MG remains the clear exception despite being evaluated on the favorable EN-derived candidate set: its much lower value of $0.32\pm0.22$ indicates that the model frequently places the ground state deep in the pool. \RR{The GSM values show a similarly compact main group, with most models falling between 1.0 and 1.4 with intervals ranging from 0.2 to 0.8 while EQ$^\ast$ is a clear redundancy outlier at a high $13.4\pm9.4$ value. The full MRR and GSM values are reported in the Summary.}}

Finally, it is instructive to compare the uMLPs against our system-specific Behler-Parrinello NN potentials fitted in previous studies for two M-Sn binaries~\cite{ak47,ak51}, with the reported total RMSE and the current ranking RMSEs of 10.2~meV/atom and 9.2~meV/atom for Li-Sn, and 9.6~meV/atom and 14.9~meV/atom for Pd-Sn, respectively. \RR{Based on the stated 45\% rate of matching the DFT global minima in our study of Li-Sn alloys~\cite{ak47}, our fitted model would have a fairly modest estimated MRR score between 0.45 and 0.73.} Nearly all uMLPs fare better, highlighting the benefit of training larger architectures with more flexible descriptors on more extensive datasets across the periodic table \RR{to capture} transferable stability trends across both structures and chemistries.

\section{Ground state perturbations}
\label{phd}

\subsection{Ground state of Zn}
\label{Zn}

\begin{figure}[t!]
    \centering
    \includegraphics[width=0.48\textwidth]{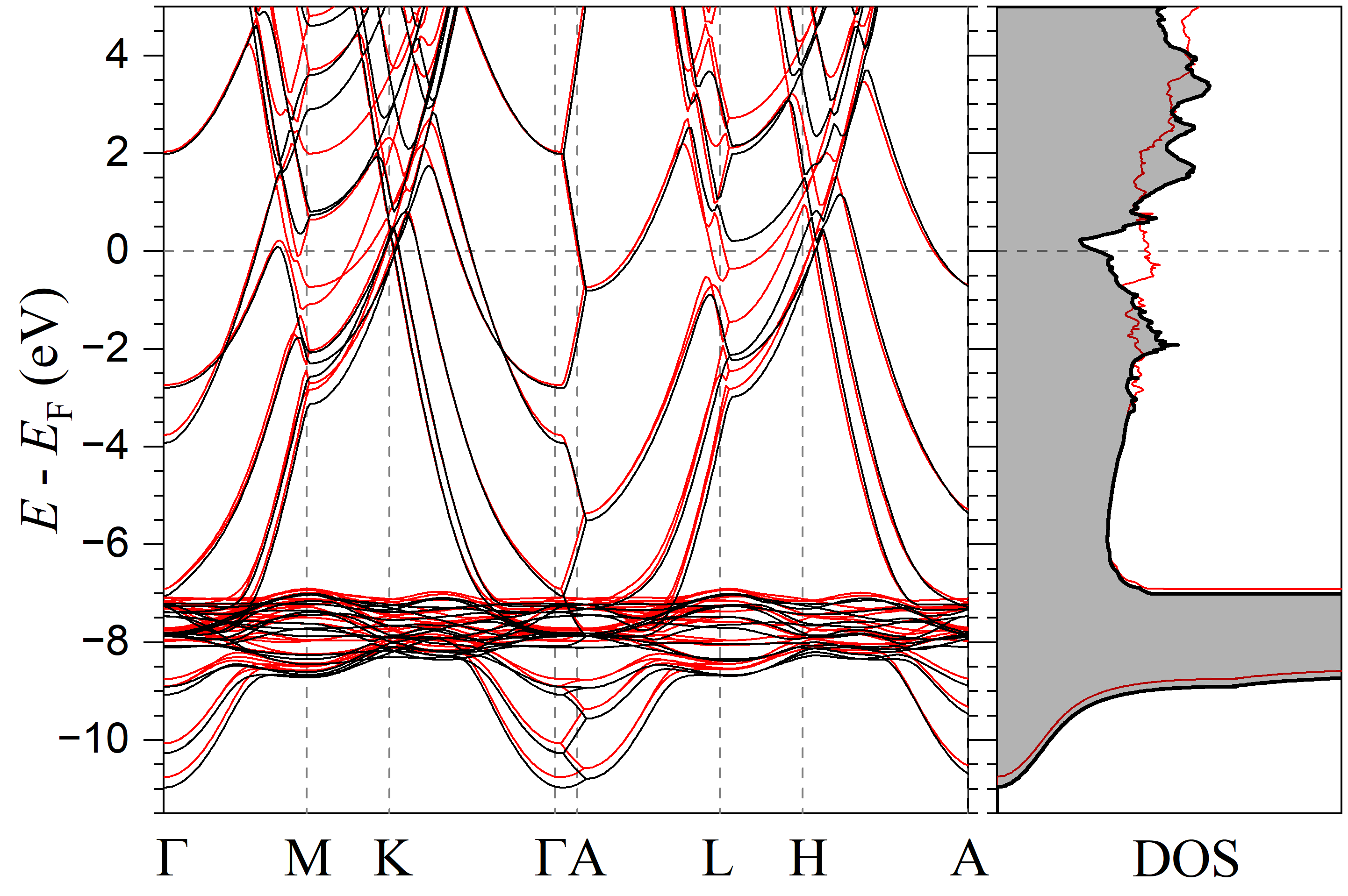}
    \caption{Band structure and density of states calculated with PBE for hcp-Zn (black) and fcc-Zn (red) represented with 6-atom hexagonal unit cells at the experimental $c/a = 1.826$.}
    \label{fig06}
\end{figure}

Among hcp metals, Zn is known to exhibit a $c/a$ ratio significantly deviating from the ideal close-packed $\sqrt{8/3}\approx 1.633$ value, an anomalous geometry attributed to the underlying electronic structure~\cite{Haussermann2001,Wedig2013,Takemura2019}. H\"aussermann and Simak~\cite{Haussermann2001} argued that the axial anisotropy is driven by a band-energy gain due to an $s$--$p$ hybridization that opens up a density of states (DOS) pseudogap and places it near $E_\textrm{F}$. The demonstrated synchronized alignment of multiple hybridization gaps from opposite directions suggests that the structural distortion is a manifestation of an electronic topological transition triggered by the varying lattice parameters. To illustrate that this stabilization is tied directly to symmetry-controlled hybridization features at $E_\textrm{F}$, we compare band structures of hcp- and fcc-Zn in Fig.~\ref{fig06}. Representation of both close-packed crystal structures with hexagonal unit cells at the experimental $c/a=1.826$ ratio extrapolated to zero Kelvin~\cite{Wedig2013} helps appreciate that it is the stacking sequence that determines the band splitting at the L and M k-points, which ultimately leads to the appearance of a pseudogap near $E_\textrm{F}$, present in hcp-Zn (ABABAB) but absent in fcc-Zn (ABCABC).

\begin{figure}[t!]
   \centering
\includegraphics[width=0.45\textwidth]{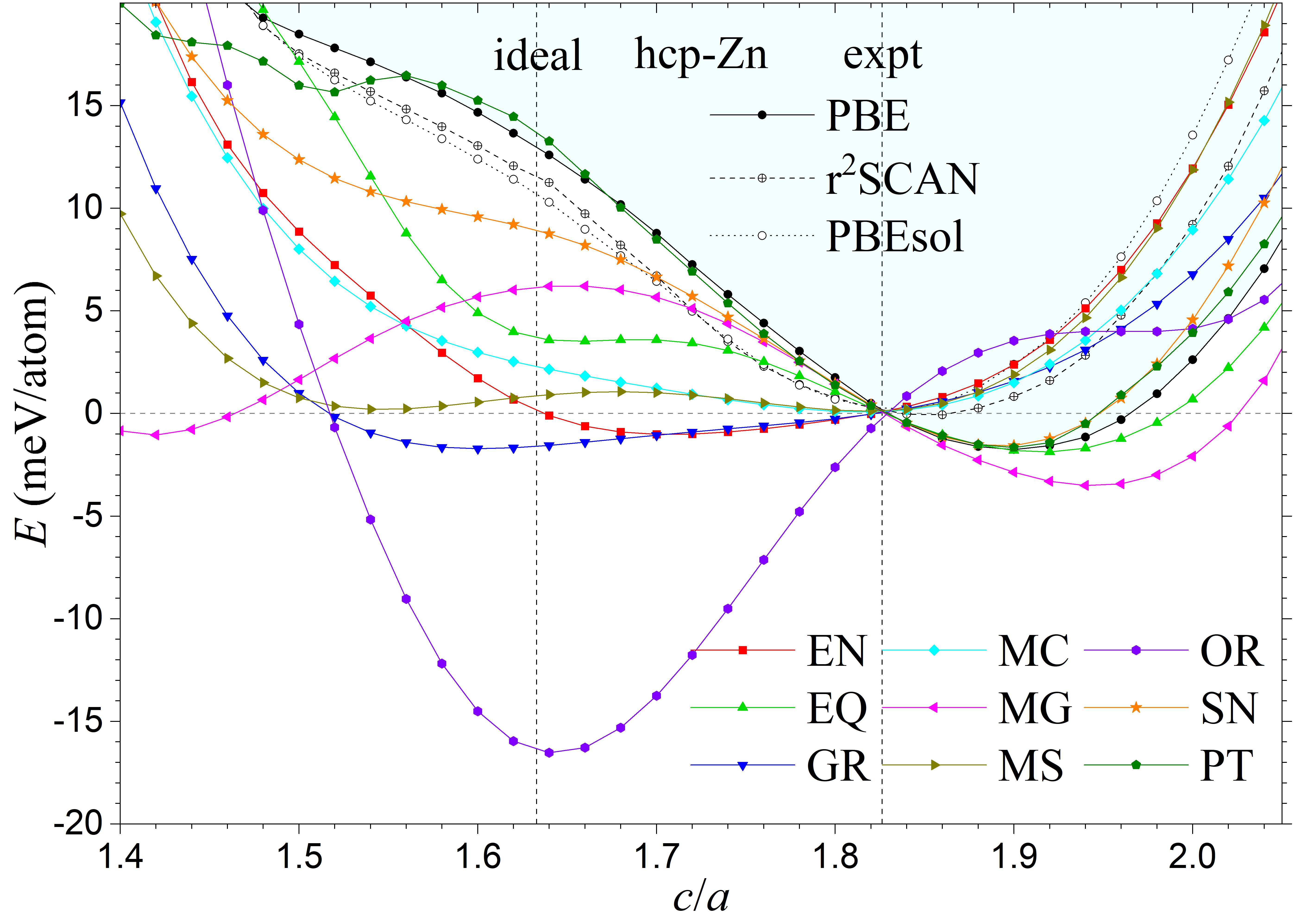}
    \caption{DFT and uMLP energy profiles of hcp-Zn as a function of $c/a$ for fixed-volume unit cells. Each curve is set to zero at the experimental $c/a = 1.826$ ratio.}
\label{coa}
\end{figure}

We examined the ability of the uMLPs to reproduce the structural anomaly in hcp-Zn by first relaxing the volume with each model and then performing single-point calculations for $c/a$ ratios around 1.826 at the fixed optimal volume. This constraint allowed us to sample the shallow trough on the PES over a wide range of structural parameters with a variation of $\sim\!20$ meV/atom, the energy scale of interest \RR{(see Fig. S7)}. Fig.~\ref{coa} shows that the three DFT approximations correctly predict the departure from the ideal packing and place the $c/a$ minimum at 1.894 (PBE), 1.826 (PBEsol), and 1.840 (r$^2$SCAN), but \RR{most} uMLPs have difficulty \RR{matching their} reference methods. \RR{SN, EQ, and PT reproduce the large $c/a$ ratio but show notable deviations from the PBE profile at lower $c/a$ values.} EN, GR, MS, \RR{and MC} are aware of the energy basin’s peculiar shape but do not properly favor the structural distortion, yielding nearly flat curves with a $\sim\! 2$ meV/atom variation between the ideal and the experimental ratios. OR fails to recognize the stabilizing effect of the distortion and defines a pronounced minimum around the ideal $c/a$. MG agrees with PBE qualitatively, predicting a large $c/a$ optimal ratio, but the feature appears fortuitous because of the presence of a second comparable minimum at unphysically small $c/a$ around 1.42. \RR{Fig. S8 illustrates a clearly better performance of the new PT, MC, and EQ parameterizations, as PT$^*$ and MC$^*$ have deep minima around the ideal $c/a$ ratio while EQ$^*$ produces $\sim\!5$ meV/atom jumps along the trajectory (the observed artifact is consistent with the findings of Fu {\it et al.}~\cite{Fu2025} who noted that grid-based spherical harmonic projections in EQ$^*$ and eSCN can introduce discretization artifacts that degrade both smoothness and energy conservation).} Our own attempts to accurately reproduce the dependence with a Behler-Parrinello NN potential constructed with our standard protocol~\cite{ak41} failed even after the distorted configurations were heavily weighted in the reference data set and the regularization factor was relaxed, with best versions (not shown) displaying EN-like behavior.

\subsection{Competing \texorpdfstring{MB$_4$}{MB4} phases for 3d metals}
\label{mb4}

\begin{figure}[t!]
   \centering
\includegraphics[width=0.48\textwidth]{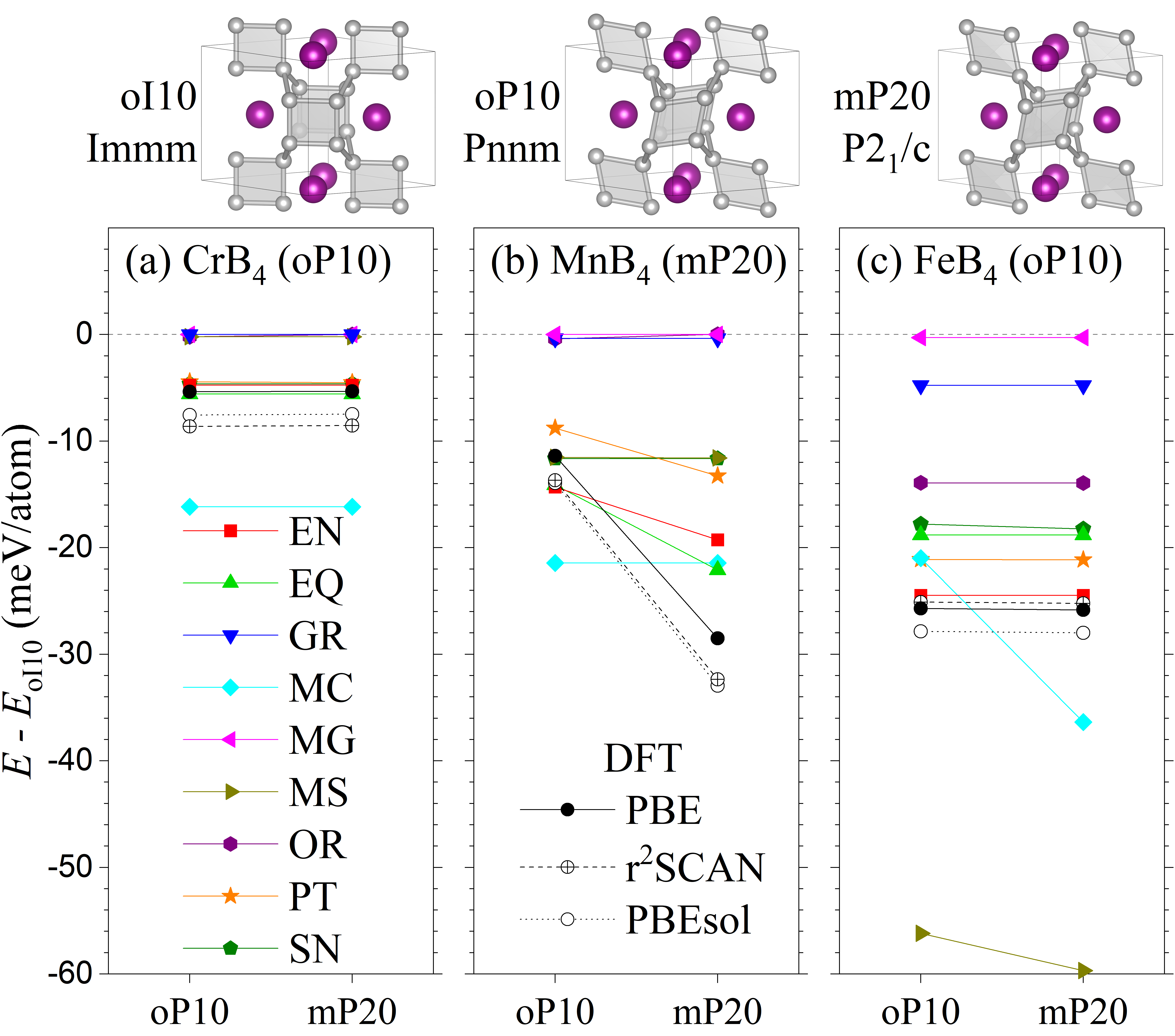}
    \caption{\label{fig-01} Stability of distorted oP10 and mP20 derivatives relative to oI10 calculated with DFT and uMLPs for MB$_4$ with (a) M = Cr, (b) M = Mn, and (c) M = Fe. The oI10 and oP10 structures are displayed with supercells to illustrate the connection with the lowest-symmetry mP20.}
\end{figure}

Ground states of 3d metal tetraborides feature a complex morphology of interlinked B$_4$ units (Fig.~\ref{fig-01}). The shape of the 3D B framework and the arrangement of the metal atoms are sensitive to the electron count and remained misidentified in CrB$_4$ and MnB$_4$ long after their discovery. In 2010, an {\it ab initio} evolutionary search for FeB$_4$ found a new oP10 structure~\cite{ak16,ak17}, a distorted derivative of the oI10 CrB$_4$ prototype, to be thermodynamically stable under moderate pressures; the phase was subsequently synthesized at pressures above 8 GPa and quenched to ambient conditions~\cite{ak26}. An {\it ab initio} reexamination of CrB$_4$ revealed that the compound also prefers the oP10 polymorph over the originally assigned oI10; the prediction was later confirmed by single-crystal XRD measurements~\cite{ak22,Knappschneider2013}. Similarly, several DFT studies reanalyzed MnB$_4$ and proposed a lower-symmetry mP20 ground state~\cite{ak28,Knappschneider2014,Gou2014}; the refined solution was verified experimentally as well~\cite{Knappschneider2014,Gou2014}. The definitive determination of ground states proved essential for this materials family, as FeB$_4$ and MnB$_4$ have been shown to superconduct at 2.7 K (1 bar)~\cite{ak26} and 14.2 K (150 GPa)~\cite{Xiang2024}, respectively. While all three compounds are non-magnetic, with ferromagnetic order vanishing in the Fe-B binary around the dilute 1:4 metal concentration~\cite{ak28}, spin fluctuations appear to play an important role in the pairing interaction, suppressing the conventional $T_{\textrm{c}}$ in FeB$_4$ and possibly contributing to the high $T_{\textrm{c}}$ in MnB$_4$. Hence, the electronically frustrated MB$_4$ compounds present a trying test case for the capability of \RR{u}MLPs to resolve competing crystal structures.

As a baseline, we examined how commonly used DFT flavors, PBE, r$^2$SCAN, and PBEsol, describe the relative stability of the three configurations. The three approximations correctly predict that CrB$_4$ and FeB$_4$ benefit from the oI10$\rightarrow$oP10 transformation that skews the B$_4$ rectangles and that MnB$_4$ further stabilizes through the oP10$\rightarrow$mP20 symmetry breaking that dimerizes Mn atoms along the metal linear chains. Fig.~\ref{fig-01} and \RR{Table S3} show that the energy gains agree to within 3--5 \RR{meV/atom: 5.4, 8.6, and 7.6 meV/atom for CrB$_4$, 25.7, 25.1, and 27.9 meV/atom for FeB$_4$, and 28.5, 32.4, and 33.0} meV/atom for MnB$_4$, respectively. In CrB$_4$, the higher stabilization found for r$^2$SCAN and PBEsol reflects the larger level of the B$_4$ distortion.

Local optimizations of the three prototypes with each uMLP were performed starting from conventional PBE-relaxed unit cells, and Fig.~\ref{fig-01} displays the energy of the oP10 and mP20 derivatives relative to the undistorted oI10 structure in each compound. For FeB$_4$, all uMLPs capture the dynamical instability of the oI10 prototype with respect to the $\Gamma$-point B phonon mode which generates oP10 with an unexpected {\it increase} of the DOS at the Fermi level~\cite{ak16}. However, MG shows no measurable energy gain, GR underestimates the stabilization by a factor of five, MS overestimates it twofold, and \RR{MC, SN, and MS} favor further monoclinic distortion. The full relaxation of the oP10 prototype reduces the right angle in the B$_4$ units to about 60$^\circ$ and increases the coordination of half of the B atoms by bringing a neighbor within a near-covalent 1.9~\AA\ bond length in all but the MG parametrization \RR{(see Table S3)}. For CrB$_4$, EQ, EN, MC, SN, PT, and MS are able to reproduce the favorability of the oP10 prototype. For MnB$_4$, GR, OR, and MG find oI10 to be locally stable, SN, MS  \RR{and MC} improve it to the oP10 derivative, \RR{and EN, EQ, and PT correctly identify mP20 as the ground state.} The three models are in quantitative agreement with PBE, matching the stabilization gain to within 15 meV/atom and the dimerized Mn-Mn nearest distances, which contract from 2.93~\AA\ in oI10 down to 2.65~\AA\ in mP20, to within 1\%. \RR{As shown in Fig. S9, all three legacy variants perform well describing CrB$_4$, but PT$^*$ and MC$^*$ fail to reproduce the mP20 stabilization for MnB$_4$ and PT$^*$ and EQ$^*$ incorrectly favor the mP20 distortion for FeB$_4$. Overall, the latest EQ and PT versions improve the description across the MB$_4$ series, whereas MC displays more pronounced deviations from the reference method.}

These tests illustrate that some of the examined uMLPs could successfully serve as surrogate models in unconstrained searches to locate the low-symmetry ground states in this materials family. The distorted prototypes would be correctly identified to have the lowest energy at the DFT level with a simple re-optimization of the candidate pool rather than an expensive phonon dispersion analysis. EQ, EN, \RR{and PT} perform best by reproducing the distorted derivatives for all three MB$_4$ compounds; SN and MC correctly identify the ground state only for CrB$_4$; GR and OR are able to recognize the favorability of the symmetry breaking only for FeB$_4$; MS points to oI10 instability across all three MB$_4$ compounds without clearly preferring any of the correct ground states; and MG predicts a very shallow basin around the oI10 parent structure in this materials family.

\subsection{Off-stoichiometry \texorpdfstring{LiB$_y$}{LiBy} phases}
\label{lib}

The structure and the composition of LiB$_{y<1}$, discovered in the late 1970s~\cite{Wang1978}, remained a puzzle for decades until an in-depth XRD analysis showed that the compound is comprised of linear B chains and a Li sublattice with incommensurate periods~\cite{Worle2000}. Subsequent modeling of LiB$_y$ with structures of varying sublattice ratios revealed that their formation energies follow a parabolic dependence on the composition~\cite{ak09}, which defines a finite range of stability around the experimentally observed $y = 0.9$ ($x=(1+y)^{-1} = 0.526$). While deviations from ideal stoichiometries are common at elevated temperatures due to the configurational entropy contribution to the formation free energy, the existence of a finite stability range at zero temperature is unique. According to the {\it ab initio} analysis~\cite{ak09}, the Li to B charge transfer does not follow a simple rigid band shift scenario because Li preferentially downshifts $p_{x,y}$ B orbitals, and the resulting DOS pseudogap moves close to the Fermi level at the observed $x$ values. The established parabolic dependence has helped explain the material’s response to H$_2$ exposure~\cite{ak13}, track its composition under compression~\cite{ak09},~\cite{Hermann2012},~\cite{ak30}, and prime its proposed function as a precursor for synthesis of a layered LiB superconductor~\cite{ak48}.

\begin{figure}[t!]
   \centering
\includegraphics[width=0.48\textwidth]{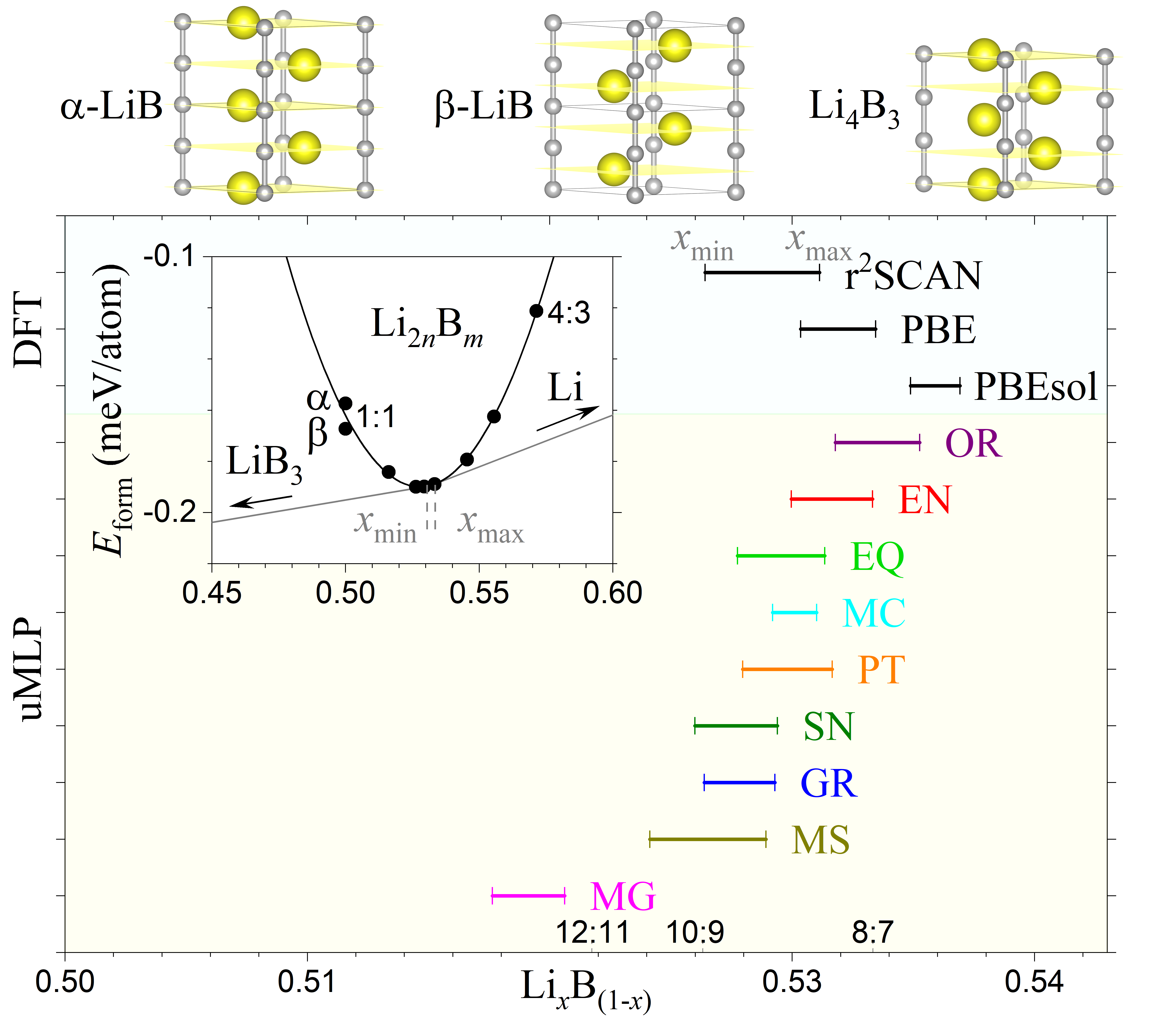}
    \caption{\label{fig-02} Stability ranges for Li$_{x}$B$_{1-x}$ phases with linear B chains for DFT and uMLPs. The inset shows the parabolic fit to the PBE formation energies, with $x_{\textrm{min}}$ and $x_{\textrm{max}}$ values determined as parabola tangents connecting LiB$_3$ and Li, respectively. }
\end{figure}

The material is a particularly valuable testbed because the off-stoichiometric ground state phases, not yet added to major materials databases, have not been included in the uMLP training sets. For example, the Materials Project has only one suboptimal representative of the material, $\alpha$-LiB. Fig.~\ref{fig-02} and \RR{Table S4} illustrate how we used the Li$_{2n}$B$_m$ series to probe whether uMLPs are capable of capturing the stability range, with $x_{\textrm{min}}$ and $x_{\textrm{max}}$ defined by tangents to the parabolic fits $\frac{1}{2}c(x-x_0)^2+E_0$ from LiB$_3$ and Li points on the formation energy plot, respectively.

Our DFT results reflect the known strong dependence of formation energies on the exchange-correlation functional: for $\alpha$-LiB and LiB$_3$, values \RR{vary up to nearly 100 meV/atom.} For this reason, our uMLP benchmarking focuses primarily on properties that are either intrinsic to the LiB$_y$ family, such as the energy difference between $\alpha$ and $\beta$ LiB phases with alternative relative placement of the B and Li sublattices, or at least invariant under affine transformations ($E'=E+A+Bx$), such as the fitted parabola curvature and the ranges of stability (see Supplementary Note I). The three DFT approximations indeed produce closely matching values for $\Delta E_{\alpha{\textrm{-LiB}}}=9.2-10.3$ meV/atom and $c=72-75$ eV/atom, but the $[x_{\textrm{min}},x_{\textrm{max}}]$ estimates show a larger dispersion, with \RR{r$^2$SCAN} and PBEsol defining non-overlapping ranges different \RR{in size} by a factor of two (Fig.~\ref{fig-02}).

The uMLPs correctly favor $\beta$ over $\alpha$ but, \RR{with the exception of GR and PT}, the energy difference is noticeably underestimated, especially for MS, SN, and \RR{PT{$^*$}}. The fitted curvatures are also consistently softer, at 60--85\% of the PBE value. The results show that \RR{the majority of} these classical models do not sufficiently penalize the sublattice shifts away from the natural Li-B registry or the compositional deviations from the optimal stoichiometry. At the same time, the PBE stability ranges are reproduced by the uMLPs with accuracy that is better (EN,  \RR{MC{$^*$}}, OR, and \RR{EQ{$^*$}}) or comparable (SN, \RR{ MC, PT, EQ, }GR, and MS) to that of the PBEsol or r$^2$SCAN estimates, while \RR{PT{$^*$}} is in reasonable agreement with the PBEsol reference. The fact that all uMLPs place the minimum of the LiB$_y$ formation energy curve in the Li-rich region, in quantitative agreement with the {\it ab initio} methods, could be an indication that the stable stoichiometries are determined predominantly by classical factors, {\it e.g.}, the ratio of the effective Li and B sizes~\cite{ak09}, and that the redistribution of electronic states is consequential rather than causal.

For assessing the quality of the LiB$_y$ description further away from equilibrium, we also calculated the relative stability between $\beta$-LiB and a layered hP8-LiB phase. The latter was predicted to stabilize under pressure~\cite{ak08}, and was later synthesized via cold compression and annealed to ambient conditions~\cite{ak30}. The conversion apparently follows a low-barrier pathway linking B chains into honeycomb B layers, a process that requires large-scale simulations to model. uMLPs could therefore shed light on the kinetics of the solid-state transformation, provided that they accurately resolve the relative energy between the hP8-LiB and $\beta$-LiB endpoints. In practice, the values vary substantially even across the DFT approximations ($-3.5$ meV/atom for PBE, $+26.6$ meV/atom for r$^2$SCAN, and $\RR{-25.6}$ meV/atom for PBEsol) and disperse over a wider range with uMLPs, with $-110$ meV/atom for \RR{MC$^*$} and $+153$ meV/atom for MG (see \RR{Table S4}). Among the considered models, only EN, \RR{MC,}  GR, OR, and \RR{PT{$^*$}} reproduce their respective references closely enough to distinguish reliably between the two competing phases with markedly different morphologies. \RR{Overall, the updated PT, MC, and EQ variants do not yield a uniform improvement in the LiB$_y$ description over their legacy counterparts (see Fig. S10).}

\begin{table*}[!t]
\caption{Aggregate performance of the considered uMLPs with respect to their reference DFT method. The models are ordered by the ranking RMSE shown in Fig.~\ref{metrics} and split into three size groups by the fitted-parameter count listed in Table~\ref{umlp}. Accuracy metrics averaged over the identified pools of low-energy candidates are shown in Fig.~\ref{metrics}. \RR{The MRR scores measure the fidelity of ranking the ground state, while the GSM reflects the degree of uMLP ground state redundancy, with both quantities defined in Section~\ref{testbed}.} Description quality of near-equilibrium configurations for selected compounds is assessed with a composite score on a 12-point scale. For hcp-Zn, the evaluation criteria are the location of the global minimum at the anomalous $c/a$ ratio, PES smoothness, and absence of pronounced secondary minima (Fig.~\ref{coa}). For the MB$_4$ family, the grades reflect the qualitative identification of the ground-state polymorph (mP20 for Mn and oP10 for Cr and Fe) and the quantitative agreement, within 10 meV/atom, of the distortion-driven stabilization (Fig.~\ref{fig-01} and Table S3). For LiB$_y$, the scores indicate the favorability of Li-rich phases ($y\approx0.9$) and reproduction of the relative stabilities among relevant LiB phases (Fig.~\ref{fig-02} and Table S4).}
\label{all}

\definecolor{topgreen}{rgb}{0.6, 0.9, 0.6}
\definecolor{midgreen}{rgb}{0.8, 1.0, 0.8}
\definecolor{warmyellow}{rgb}{1.0, 0.95, 0.7}
\definecolor{softred}{rgb}{1.0, 0.8, 0.8}

\newcommand{\dg}{\cellcolor{topgreen}}
\newcommand{\lgc}{\cellcolor{midgreen}}
\newcommand{\yc}{\cellcolor{warmyellow}}
\newcommand{\rc}{\cellcolor{softred}}

\begin{ruledtabular}
\small
\renewcommand{\arraystretch}{1.1}
\begin{tabular}{l c w{c}{1.35cm} *{3}{w{c}{1.2cm}} w{c}{2.1cm} w{c}{1.9cm}*{3}{w{c}{1.2cm}}}
\multicolumn{3}{c}{uMLP}
& \multicolumn{5}{c}{\RR{Merged pool metrics}}
& \multicolumn{3}{c}{Ground state perturbations} \\
\cline{1-3} \cline{4-8} \cline{9-11}
\rule{0pt}{3ex}
Name & ID & Size & RMSE & Prox(E) & Prox(S) & \RR{MRR} & \RR{GSM}
& hcp-Zn & MB$_4$ & LiB$_y$ \\
\hline

\RR{EquiformerV3}
& EQ\phantom{{$^*$}}
& Medium
&  3.41
&  0.33
&  0.76
&  0.91~$\pm$~0.18
&  1.1~$\pm$~0.4
&  11
&  12
&  8 \\

EquiformerV2
& \RR{EQ{$^{*}$}}
& Large
&  4.22
&  1.17
&  0.69
&  0.82~$\pm$ 0.24
&  13.4~$\pm$~9.4\phantom{{$0$}}
&  6
&  12
&  7 \\

\RR{PET-OAM}
& PT\phantom{{$^*$}}
& Large
&  4.31
&  0.51
&  0.76
&  0.91~$\pm$~0.18
&  1.2~$\pm$~0.3
&  11
&  10
&  8 \\

eSEN
& EN\phantom{{$^*$}}
& Medium
&  4.49
&  0.41
&  0.75
&  0.91~$\pm$~0.18
&  1.1~$\pm$~0.2
&  8
&  12
&  11 \\

\RR{MACE-mh-1}
& MC\phantom{{$^*$}}
& Medium
&  9.17
&  2.91
&  0.66
&  0.84~$\pm$~0.21
&  1.1~$\pm$~0.2
&  9
&  4
&  11 \\

Orb-v3
& OR\phantom{{$^*$}}
& Medium
&  9.64
&  2.45
&  0.66
&  0.91~$\pm$~0.18
&  1.1~$\pm$~0.2
&  2
&  2
&  11 \\

SevenNet
& SN\phantom{{$^*$}}
& Medium
&  9.92
&  2.90
&  0.63
&  0.91~$\pm$~0.18
&  1.4~$\pm$~0.8
&  12
&  6
&  6 \\

GRACE
& GR\phantom{{$^*$}}
& Medium
&  11.68
&  2.29
&  0.63
&  0.71~$\pm$~0.25
&  1.1~$\pm$~0.2
&  8
&  2
&  12 \\

PET-MAD
& \RR{PT{$^{*}$}}
& Small
&  15.94
&  6.50
&  0.54
&  0.91~$\pm$~0.18
&  1.4~$\pm$~0.4
&  4
&  6
&  10 \\

MACE-MATPES
& \RR{MC{$^*$}}
& Medium
&  18.28
&  5.99
&  0.55
&  0.70~$\pm$~0.25
&  1.1~$\pm$~0.2
&  4
&  6
&  9 \\

MatterSim
& MS\phantom{{$^*$}}
& Small
&  21.66
&  7.96
&  0.56
&  0.86~$\pm$~0.19
&  1.0~$\pm$~0.0
&  6
&  0
&  6 \\

M3GNet
& MG\phantom{{$^*$}}
& Small
&  \phantom{{$\daggerfootnotemark$}}22.45\RR{\daggerfootnotemark}
&  \phantom{{$\daggerfootnotemark$}}65.0\RR{\daggerfootnotemark}
&  \phantom{{$\daggerfootnotemark$}}0.44\RR{\daggerfootnotemark}
&  0.32\RR{\daggerfootnotemark}~$\pm$~0.22\phantom{{$\daggerfootnotemark$}}
&  1.1\RR{\daggerfootnotemark}~$\pm$~0.2\phantom{{$\daggerfootnotemark$}}
&  0
&  0
&  6 \\

\end{tabular}
\end{ruledtabular}

\daggerfootnotetext{\RR{Metrics evaluated with EN curated pool as discussed in Section~\ref{evos}.}}

\end{table*}

\section{Summary}
\label{sum}

This study presents a systematic benchmark of nine state-of-the-art uMLPs \RR{and three legacy models} through evolutionary crystal structure prediction. Unlike traditional testing frameworks that often rely on libraries of commonly observed high-symmetry structures, our protocol probed the models’ readiness for practical unconstrained explorations that have a track record in the accelerated discovery of new thermodynamically stable prototypes~\cite{ak47,ak51}. By focusing on multicomponent materials, we evaluated the uMLPs' ability to describe diverse bonding mechanisms and complex ground states. The inclusion of three compounds with recently proposed ground states, Li$_3$Sn, Pd$_5$Sn$_3$, and MgB$_3$C$_3$, further tested whether out-of-the-box uMLPs can handle configurations absent from their training sets.

The results reveal an impressive level of predictive maturity across \RR{modern uMLPs}. With the exception of the earliest architecture \RR{(MG)}, the considered uMLPs successfully managed unconstrained searches, locating ground states for most compounds (see Table~\ref{main} and Fig. \ref{metrics}). The remarkably low rate of unphysical artifacts indicates that current training sets provide a robust baseline for inorganic chemistry. AgClO$_4$ was the sole compound where all models struggled\RR{, likely} due to insufficient sampling of specific oxygen bonding environments.

\RR{Our combined analysis of over three million uMLP local relaxations and more than ten thousand DFT calculations summarized in Table~\ref{all} allows for a more contextual understanding of model behavior that a single accuracy metric cannot capture alone. For example, EQ{$^*$} has the second-lowest ranking RMSE, yet it generated numerous redundant candidates with an average GSM of $13.4\pm9.4$, and, therefore, has limited practical value for evolutionary searches. At the other end of the accuracy spectrum, MG exhibits RMSE values typical of smaller uMLPs and comparable to the DFT systematic spread, but proves to be singularly ineffective in nearly all unconstrained runs.}

Rather unexpectedly, the extensive structure searches carried out purely for benchmarking purposes identified three phases lying below the previously reported global minima at the reference DFT level. The tI10-Na$_2$CN$_2$ candidate is favored only by PBE, \RR{whereas the mS24-CYI candidate is stabilized by PBE and r$^2$SCAN but not by PBEsol or vdW-corrected treatments, making both} likely approximation-\RR{dependent} artifacts. The oI28-MgB$_3$C$_3$ phase ranks below the proposed layered superconducting polymorph~\cite{ak54} across all considered exchange-correlation functionals, indicating that deintercalation of the MgB$_2$C$_2$ precursor may lead to products with morphologies unrelated to the starting honeycomb framework.

We also examined whether current uMLPs can map subtle near-equilibrium PES features arising from electronic structure peculiarities. As in our multi-metric evaluation of global optimization performance, we assigned composite qualitative and quantitative scores to assess reproduction of the relevant observables described in the caption of Table~\ref{all}. In this case, we focused on prospective simulations probing structural response and phase transformations on scales for which direct DFT validation may be prohibitively expensive.

For the well-studied hcp-Zn, most uMLPs struggled to reproduce the shallow basin around the anomalous $c/a$ ratio defined by the electronic band topology~\cite{Haussermann2001}. Only SN\RR{, EQ, and PT} closely follow the reference energy profile and could be used in \RR{MD} simulations to model, in particular, the observed $c/a$ response to temperature~\cite{Wedig2013}. For the CrB$_4$, FeB$_4$, and MnB$_4$ compounds, several uMLPs correctly reproduce the chemistry-dependent symmetry lowering among the competing oI10, oP10, and mP20 polymorphs. The performance is encouraging for future large-scale exploration of the (Cr, Mn, Fe)B$_4$ phase space aimed at identifying synthesizable superconducting and superhard materials without exhaustive {\it ab initio} stability analyses for every distortion pattern~\cite{ak16,ak17,ak28}. For off-stoichiometric LiB$_y$, all uMLPs correctly place the stability minimum in the Li-rich region near $y\approx0.9$, despite standard databases still listing LiB as the ground-state composition. Detailed modeling of the experimentally observed structural transformation from the ambient-pressure precursor to a layered phase with potential for high-$T_\textrm{c}$ superconductivity~\cite{ak30,ak48} demands accurate description of the competing Li-B phases. While several uMLPs agree with the reference method, the common semi-local DFT approximations themselves do not account for the vdW interactions important in this materials class.

Overall, the benchmark indicates that current pretrained uMLPs can already boost {\it ab initio} global optimization \RR{of nonmagnetic inorganic crystal structures under ambient pressure} and accurately describe selected near-equilibrium PES features, with \RR{EQ, PT}, and EN delivering the most consistent performance across the tested cases. The results are consistent with the general trend that larger models with more expressive descriptors and broader training coverage transfer more reliably across competing structures and chemistries. In this respect, system-specific potentials may no longer need to be the default starting point for crystal structure exploration. Any application to a specific materials problem should still be preceded by case-specific validation and, when needed, retraining against an appropriate reference set. This caution is especially important because no exchange-correlation approximation is universally accurate across the periodic table.

\section*{Conflicts of interest}
\label{con}

There are no conflicts to declare.

\section*{Data Availability}
\label{data-avail}
The data supporting the findings of this study \RR{and descriptions of our workflows (Supplementary Note II)} are available within the \RR{Supplemental Material} \RR{and in the Zenodo repository at 10.5281/zenodo.20880271.} \RR{The open-source MAISE code can be acquired at https://github.com/maise-guide/maise.} Additional data are available from the corresponding author upon request.

Supplementary Material:
\RR{uMLP single-core CPU efficiency tests,} evolutionary search and native minima pool details, \RR{merged pool details,} representative AgClO$_4$ artifact, \RR{performance metrics for legacy models, ranking RMSEs per compound for merged pools,} violin plots for pool-based energy differences between uMLPs and DFT, \RR{native and merged pool ranking RMSE comparison,} PBE \RR{PES} for hcp-Zn, \RR{ground state perturbation tests for legacy models,} stability and structural information of MB$_4$ phases (M = Cr, Mn, and Fe), stability of Li\RR{$_x$}B\RR{$_{1-x}$} phases, LiB$_y$ properties under affine transformations, \RR{example MAISE settings for evolutionary searches, BLOCK file for MgB$_3$C$_3$ TETRIS generation, evolutionary optimization and perturbation calculation workflows, information for used VASP pseudopotentials,} structural information for select phases in CIF format.

\begin{acknowledgments} 
The authors thank Aidan Thorn and Erin Delargy for their help with constructing neural network potentials for Zn and acknowledge support from the National Science Foundation (NSF) (Award No. DMR-2320073). This work used the Expanse system at the San Diego Supercomputer Center via allocation TG-DMR180071. Expanse is supported by the Extreme Science and Engineering Discovery Environment (XSEDE) program~\cite{XSEDE} through NSF Award No. ACI-1548562. \RR{The work of O.G. was supported by Grant \#13001218-080 through SCALE: A Microelectronic Workforce Development Production Project.}

\end{acknowledgments}

\bibliography{refs}

@article{ak41,
  author  = {Hajinazar, Samad and Thorn, Aidan and Sandoval, Ernesto D. and Kharabadze, Saba and Kolmogorov, Aleksey N.},
  title   = {{{MAISE}}: {Construction} of neural network interatomic models and evolutionary structure optimization},
  journal = {Comput. Phys. Commun.},
  volume  = {259},
  pages   = {107679},
  year    = {2021},
  doi     = {10.1016/j.cpc.2020.107679}
}

@article{ak40,
  author  = {Thorn, Aidan and Rojas-Nunez, Javier and Hajinazar, Samad and Baltazar, Samuel E. and Kolmogorov, Aleksey N.},
  title   = {Toward {{\it ab Initio} Ground States} of {Gold Clusters} via {Neural Network Modeling}},
  journal = {J. Phys. Chem. C},
  volume  = {123},
  pages   = {30088--30098},
  year    = {2019},
  doi     = {10.1021/acs.jpcc.9b08517}
}

@article{ak51,
  author  = {Thorn, Aidan and Gochitashvili, Daviti and Kharabadze, Saba and Kolmogorov, Aleksey N.},
  title   = {Machine learning search for stable binary {Sn} alloys with {Na}, {Ca}, {Cu}, {Pd}, and {Ag}},
  journal = {Phys. Chem. Chem. Phys.},
  volume  = {25},
  pages   = {22415--22436},
  year    = {2023},
  doi     = {10.1039/D3CP02817H}
}

@article{ak47,
  author  = {Kharabadze, Saba and Thorn, Aidan and Koulakova, Ekaterina A. and Kolmogorov, Aleksey N.},
  title   = {Prediction of stable {Li-Sn} compounds: boosting ab initio searches with neural network potentials},
  journal = {npj Comput. Mater.},
  volume  = {8},
  pages   = {136},
  year    = {2022},
  doi     = {10.1038/s41524-022-00825-4}
}

@article{ak08,
  author  = {Kolmogorov, Aleksey N. and Curtarolo, Stefano},
  title   = {Prediction of different crystal structure phases in metal borides: {A} lithium monoboride analog to {{MgB}$_2$}},
  journal = {Phys. Rev. B},
  volume  = {73},
  pages   = {180501},
  year    = {2006},
  doi     = {10.1103/PhysRevB.73.180501}
}

@article{ak16,
  author  = {Kolmogorov,  A. N. and Shah,  S. and Margine,  E. R. and Bialon,  A. F. and Hammerschmidt,  T. and Drautz,  R.},
  title   = {New {Superconducting} and {Semiconducting} {Fe-B} {Compounds Predicted} with an {{\it Ab Initio} Evolutionary Search}},
  journal = {Phys. Rev. Lett.},
  volume  = {105},
  pages   = {217003},
  year    = {2010},
  doi     = {10.1103/PhysRevLett.105.217003}
}

@article{ak17,
  author  = {Bialon,  A. F. and Hammerschmidt,  T. and Drautz,  R. and Shah,  S. and Margine,  E. R. and Kolmogorov,  A. N.},
  title   = {Possible routes for synthesis of new boron-rich {{Fe--B}} and {{Fe}$_{1-x}${Cr}$_x${B}$_4$} compounds},
  journal = {Appl. Phys. Lett.},
  volume  = {98},
  pages   = {081901},
  year    = {2011},
  doi     = {10.1063/1.3556564}
}

@article{ak28,
title = {{Stability of 41 metal–boron systems at 0GPa and 30GPa from first principles}},
journal = {Calphad},
volume = {46},
pages = {184-204},
year = {2014},
issn = {0364-5916},
doi = {https://doi.org/10.1016/j.calphad.2014.03.005},
url = {https://www.sciencedirect.com/science/article/pii/S0364591614000364},
author = {A.G. {Van Der Geest} and A.N. Kolmogorov}
}

@article{XSEDE,
 author = {J. Towns and T. Cockerill and M. Dahan and I. Foster and K. Gaither and A. Grimshaw and V. Hazlewood and S. Lathrop and D. Lifka and G. D. Peterson and R. Roskies and J. R. Scott and N. Wilkins-Diehr},
 title = {{XSEDE}: Accelerating scientific discovery},
 journal = {Comput. Sci. {\&} Eng.},
 volume = {16},
 pages = {62-74},
 year = {2014},
 doi={10.1109/MCSE.2014.80},
 url = {https://ieeexplore.ieee.org/abstract/document/6866038}
}

@article{ak21,
  title = {{Spin Waves and Revised Crystal Structure of Honeycomb Iridate ${\mathrm{Na}}_{2}{\mathrm{IrO}}_{3}$}},
  author = {Choi, S. K. and Coldea, R. and Kolmogorov, A. N. and Lancaster, T. and Mazin, I. I. and Blundell, S. J. and Radaelli, P. G. and Singh, Yogesh and Gegenwart, P. and Choi, K. R. and Cheong, S.-W. and Baker, P. J. and Stock, C. and Taylor, J.},
  journal = {Phys. Rev. Lett.},
  volume = {108},
  issue = {12},
  pages = {127204},
  numpages = {5},
  year = {2012},
  month = {Mar},
  publisher = {American Physical Society},
  doi = {10.1103/PhysRevLett.108.127204},
  url = {https://link.aps.org/doi/10.1103/PhysRevLett.108.127204}
}

@article{PBE,
  title={Generalized gradient approximation made simple},
  author={Perdew, J. P. and Burke, K. and Ernzerhof, M.},
  journal={Phys. Rev. Lett.},
  volume={77},
  pages={3865},
  year={1996},
  url={https://journals.aps.org/prl/abstract/10.1103/PhysRevLett.77.3865}
}

@article{PBEsol,
  title = {Restoring the Density-Gradient Expansion for Exchange in Solids and Surfaces},
  author = {Perdew, John P. and Ruzsinszky, Adrienn and Csonka, G\'abor I. and Vydrov, Oleg A. and Scuseria, Gustavo E. and Constantin, Lucian A. and Zhou, Xiaolan and Burke, Kieron},
  journal = {Phys. Rev. Lett.},
  volume = {100},
  issue = {13},
  pages = {136406},
  numpages = {4},
  year = {2008},
  month = {Apr},
  publisher = {American Physical Society},
  doi = {10.1103/PhysRevLett.100.136406},
  url = {https://link.aps.org/doi/10.1103/PhysRevLett.100.136406}
}

@article{Klime2011,
  title={Van der {Waals} density functionals applied to solids},
  author={Klime{\v{s}}, J. and Bowler, D. R. and Michaelides, A.},
  journal={Phys. Rev. B},
  volume={83},
  pages={195131},
  year={2011},
  url={https://journals.aps.org/prb/abstract/10.1103/PhysRevB.83.195131}
}

@article{Monkhorst1976,
  title={Special points for {Brillouin}-zone integrations},
  author={Monkhorst, H. J. and Pack, J. D.},
  journal={Phys. Rev. B},
  volume={13},
  pages={5188},
  year={1976},
  url={https://journals.aps.org/prb/abstract/10.1103/PhysRevB.13.5188}
}

@article{Kresse1996A,
  title={{Efficient iterative schemes for $ab~initio$ total-energy calculations using a plane-wave basis set}},
  author={Kresse, G. and Furthm{\"u}ller, J.},
  journal={Phys. Rev. B},
  volume={54},
  number={16},
  pages={11169},
  year={1996},
  publisher={APS},
  url={https://doi.org/10.1103/PhysRevB.54.11169}
}

@article{Blochl1994A,
  title={Projector augmented-wave method},
  author={Bl{\"o}chl, P. E.},
  journal={Phys. Rev. B},
  volume={50},
  number={24},
  pages={17953},
  year={1994},
  publisher={APS},
  url={https://doi.org/10.1103/PhysRevB.50.17953}
}

@article{Behler2007,
  title = {Generalized Neural-Network Representation of High-Dimensional Potential-Energy Surfaces},
  author = {Behler, J\"org and Parrinello, Michele},
  journal = {Phys. Rev. Lett.},
  volume = {98},
  issue = {14},
  pages = {146401},
  numpages = {4},
  year = {2007},
  month = {Apr},
  publisher = {American Physical Society},
  doi = {10.1103/PhysRevLett.98.146401},
  url = {https://link.aps.org/doi/10.1103/PhysRevLett.98.146401}
}

@article{Pickard2022,
  title = {{Ephemeral data derived potentials for random structure search}},
  author = {Pickard, Chris J.},
  journal = {Phys. Rev. B},
  volume = {106},
  issue = {1},
  pages = {014102},
  numpages = {15},
  year = {2022},
  month = {Jul},
  publisher = {American Physical Society},
  doi = {10.1103/PhysRevB.106.014102},
  url = {https://link.aps.org/doi/10.1103/PhysRevB.106.014102}
}

@article{ak54,
  author  = {Tomassetti, Charlsey R. and Gochitashvili, Daviti and Renskers, Christopher and Margine, Elena R. and Kolmogorov, Aleksey N.},
  title   = {First-principles design of ambient-pressure {{Mg}$_x${B}$_2${C}$_2$} and {{Na}$_x$BC} superconductors},
  journal = {Phys. Rev. Mater.},
  volume  = {8},
  number  = {11},
  pages   = {114801},
  year    = {2024},
  doi     = {10.1103/PhysRevMaterials.8.114801}
}

@article{Riebesell2025,
  author  = {Riebesell, Janosh and Goodall, Rhys E. A. and Benner, Philipp and Chiang, Yuan and Deng, Bowen and Ceder, Gerbrand and Asta, Mark and Lee, Alpha A. and Jain, Anubhav and Persson, Kristin A.},
  title   = {A framework to evaluate machine learning crystal stability predictions},
  journal = {Nat. Mach. Intell.},
  volume  = {7},
  number  = {6},
  pages   = {836--847},
  year    = {2025},
  doi     = {10.1038/s42256-025-01055-1}
}

@article{Peng2026,
  author  = {Peng, Anyang and Cai, Chun and Guo, Mingyu and Zhang, Duo and Zhang, Chengqian and Jiang, Wanrun and Wang, Yinan and Loew, Antoine and Wu, Chengkun and E, Weinan and Zhang, Linfeng and Wang, Han},
  title   = {LAMBench: a benchmark for large atomistic models},
  journal = {npj Comput. Mater.},
  year    = {2026},
  volume  = {12},
  number  = {1},
  pages   = {62},
  doi     = {10.1038/s41524-025-01929-3}
}

@article{Yu2024,
  author  = {Yu, Haochen and Giantomassi, Matteo and Materzanini, Giuliana and Wang, Junjie and Rignanese, Gian-Marco},
  title   = {Systematic assessment of various universal machine-learning interatomic potentials},
  journal = {Mater. Genome Eng. Adv.},
  volume  = {2},
  pages   = {e58},
  year    = {2024},
  doi     = {10.1002/mgea.58}
}

@article{Chiang2025,
  author  = {Chiang, Yuan and Kreiman, Tobias and Zhang, Christine and Kuner, Matthew C. and Weaver, Elizabeth and Amin, Ishan and Park, Hyunsoo and Lim, Yunsung and Kim, Jihan and Chrzan, Daryl and Walsh, Aron and Blau, Samuel M. and Asta, Mark and Krishnapriyan, Aditi S.},
  title   = {{{MLIP} Arena: Advancing Fairness and Transparency in Machine Learning Interatomic Potentials via an Open, Accessible Benchmark Platform}},
  journal = {arXiv preprint arXiv:2509.20630},
  year    = {2025},
  doi     = {10.48550/arXiv.2509.20630}
}

@article{Lyngby2024,
  title = {Bayesian optimization of atomic structures with prior probabilities from universal interatomic potentials},
  volume = {8},
  ISSN = {2475-9953},
  url = {http://dx.doi.org/10.1103/PhysRevMaterials.8.123802},
  DOI = {10.1103/physrevmaterials.8.123802},
  number = {12},
  journal = {Phys. Rev. Mater.},
  publisher = {American Physical Society (APS)},
  author = {Lyngby,  Peder and Larsen,  Casper and Jacobsen,  Karsten Wedel},
  year = {2024},
  month = Dec 
}

@article{Loew2025B,
  title = {Universal machine learning potentials under pressure},
  volume = {9},
  ISSN = {2515-7639},
  url = {http://dx.doi.org/10.1088/2515-7639/ae2ba8},
  DOI = {10.1088/2515-7639/ae2ba8},
  number = {1},
  journal = {Journal of Physics: Materials},
  publisher = {IOP Publishing},
  author = {Loew,  Antoine and Schmidt,  Jonathan and Botti,  Silvana and Marques,  Miguel A L},
  year = {2025},
  month = dec,
  pages = {015010}
}

@article{Tahmasbi2025,
  author  = {Tahmasbi, Hossein and Kn{\"{u}}pfer, Andreas and K{\"{u}}hne, Thomas D. and Mirhosseini, Hossein},
  title   = {Benchmarking {Universal} {Machine} {Learning} {Interatomic} {Potentials} on {Elemental} {Systems}},
  journal = {arXiv preprint arXiv:2512.20230},
  year    = {2025},
  doi     = {10.48550/arXiv.2512.20230}
}

@article{Loew2025,
  title   = {Universal machine learning interatomic potentials are ready for phonons},
  author  = {Loew, Antoine and Sun, Dewen and Wang, Hai-Chen and Botti, Silvana and Marques, Miguel A. L.},
  journal = {npj Comput. Mater.},
  volume  = {11},
  pages   = {10},
  year    = {2025},
  doi     = {10.1038/s41524-025-01650-1}
}

@article{Chen2022,
  title = {A universal graph deep learning interatomic potential for the periodic table},
  author = {Chen, Chi and Ong, Shyue Ping},
  journal = {Nat. Comput. Sci.},
  volume = {2},
  number = {11},
  pages = {718--728},
  year = {2022},
  doi = {10.1038/s43588-022-00349-3}
}

@article{Ko2025,
  author  = {Ko, Tsz Wai and Deng, Bowen and Nassar, Marcel and Barroso-Luque, Luis and Liu, Runze and Qi, Ji and Thakur, Atul C. and Mishra, Adesh Rohan and Liu, Elliott and Ceder, Gerbrand and Miret, Santiago and Ong, Shyue Ping},
  title   = {{Materials Graph Library (MatGL)}, an open-source graph deep learning library for materials science and chemistry},
  journal = {npj Comput. Mater.},
  volume  = {11},
  pages   = {253},
  year    = {2025},
  doi     = {10.1038/s41524-025-01742-y}
}

@article{Hjorth2017,
  author  = {Larsen, Ask Hjorth and Mortensen, Jens J{\o}rgen and Blomqvist, Jakob and Castelli, Ivano Eligio and Christensen, Rune and Du{\l}ak, Marcin and Friis, Jesper and Groves, Michael and Hammer, Bj{\o}rk and Hargus, Cory and Hermes, Eric and Jennings, Paul C. and Jensen, Peter Bjerre and Kermode, James and Kitchin, John R. and Kolsbjerg, Esben and Kubal, Joseph and Kaasbjerg, Kristen and Lysgaard, Steen and Maronsson, Jon Bergmann and Maxson, Tristan and Olsen, Thomas and Pastewka, Lars and Peterson, Andrew T. and Rostgaard, Carsten and Schi{\o}tz, Jakob and Sch{\"{u}}tt, Ole and Strange, Mikkel and Thygesen, Kristian Sommer and Vegge, Tejs and Vilhelmsen, Lasse and Walter, Michael and Zeng, Zhenhua and Jacobsen, Karsten Wedel},
  title   = {The {Atomic Simulation Environment} --- {A} {Python} library for working with atoms},
  journal = {J. Phys. Condens. Matter},
  volume  = {29},
  number  = {27},
  pages   = {273002},
  year    = {2017},
  doi     = {10.1088/1361-648X/aa680e}
}

@article{Mair1999,
  author  = {Mair, Gunther and von Schnering, Hans-Georg and W{\"{o}}rle, Michael and Nesper, Reinhard},
  title   = {{Dilithium} {Hexaboride}, {{Li}$_2${B}$_6$}},
  journal = {Z. Anorg. Allg. Chem.},
  volume  = {625},
  number  = {7},
  pages   = {1207--1211},
  year    = {1999},
  doi     = {10.1002/(SICI)1521-3749(199907)625:7<1207::AID-ZAAC1207>3.0.CO;2-9}
}

@article{Becher1962,
  title = {Darstellung und {Struktur} des {Berylliumborids} {{Be}$_4${B}}},
  author = {Becher, H. J. and Sch{\"{a}}fer, A.},
  journal = {Z. Anorg. Allg. Chem.},
  volume = {318},
  number = {5--6},
  pages = {304--312},
  year = {1962},
  doi = {10.1002/zaac.19623180509}
}

@article{Howard1991,
  author  = {Howard, C. J. and Sabine, T. M. and Dickson, F.},
  title   = {Structural and thermal parameters for rutile and anatase},
  journal = {Acta Crystallogr. Sect. B: Struct. Sci.},
  volume  = {47},
  number  = {4},
  pages   = {462--468},
  year    = {1991},
  doi     = {10.1107/S010876819100335X}
}

@article{Eguchi2011,
  author  = {Eguchi, G. and Peets, D. C. and Kriener, M. and Maeno, Y. and Nishibori, E. and Kumazawa, Y. and Banno, K. and Maki, S. and Sawa, H.},
  title   = {Crystallographic and superconducting properties of the fully gapped noncentrosymmetric $5d$-electron superconductors {Ca$M$Si$_3$} (${M} = \text{Ir}$, \text{Pt})},
  journal = {Phys. Rev. B},
  volume  = {83},
  pages   = {024512},
  year    = {2011},
  doi     = {10.1103/PhysRevB.83.024512}
}

@inproceedings{Batatia2022,
  author    = {Batatia, Ilyes and Kov{\'{a}}cs, D{\'{a}}vid P{\'{e}}ter and Simm, Gregor N. C. and Ortner, Christoph and Cs{\'{a}}nyi, G{\'{a}}bor},
  title     = {{{MACE}}: {Higher} {Order} {Equivariant} {Message} {Passing} {Neural} {Networks} for {Fast} and {Accurate} {Force} {Fields}},
  booktitle = {Adv. Neural Inf. Process. Syst. (NeurIPS)},
  volume    = {35},
  pages     = {11423--11436},
  year      = {2022},
  doi       = {10.48550/arXiv.2206.07697}
}

@article{Batatia2025,
  title = {The design space of {E}(3)-equivariant atom-centred interatomic potentials},
  volume = {7},
  ISSN = {2522-5839},
  url = {http://dx.doi.org/10.1038/s42256-024-00956-x},
  DOI = {10.1038/s42256-024-00956-x},
  number = {1},
  journal = {Nature Machine Intelligence},
  publisher = {Springer Science and Business Media LLC},
  author = {Batatia,  Ilyes and Batzner,  Simon and Kovács,  Dávid Péter and Musaelian,  Albert and Simm,  Gregor N. C. and Drautz,  Ralf and Ortner,  Christoph and Kozinsky,  Boris and Csányi,  Gábor},
  year = {2025},
  month = jan,
  pages = {56–67}
}

@article{Batatia2025b,
  doi = {10.48550/arXiv.2510.25380},
  url = {https://arxiv.org/abs/2510.25380},
  author = {Batatia,  Ilyes and Lin,  Chen and Hart,  Joseph and Kasoar,  Elliott and Elena,  Alin M. and Norwood,  Sam Walton and Wolf,  Thomas and Csányi,  Gábor},
  title = {Cross Learning between Electronic Structure Theories for Unifying Molecular,  Surface,  and Inorganic Crystal Foundation Force Fields},
  journal = {arXiv preprint arXiv:2510.25380},
  year = {2025},
  copyright = {arXiv.org perpetual,  non-exclusive license}
}

@article{Yang2024,
  author  = {Yang, Han and Hu, Chenxi and Zhou, Yichi and Liu, Xixian and Shi, Yu and Li, Jielan and Li, Guanzhi and Chen, Zekun and Chen, Shuizhou and Zeni, Claudio and Horton, Matthew and Pinsler, Robert and Fowler, Andrew and Z{\"{u}}gner, Daniel and Xie, Tian and Smith, Jake and Sun, Lixin and Wang, Qian and Kong, Lingyu and Liu, Chang and Hao, Hongxia and Lu, Ziheng},
  title   = {{{MatterSim}}: {A} {Deep} {Learning} {Atomistic} {Model} {Across} {Elements}, {Temperatures} and {Pressures}},
  journal = {arXiv preprint arXiv:2405.04967},
  year    = {2024},
  doi     = {10.48550/arXiv.2405.04967}
}

@article{Neumann2024,
  author  = {Neumann, Mark and Gin, James and Rhodes, Benjamin and Bennett, Steven and Li, Zhiyi and Choubisa, Hitarth and Hussey, Arthur and Godwin, Jonathan},
  title   = {{{Orb}}: {A} {Fast}, {Scalable} {Neural} {Network} {Potential}},
  journal = {arXiv preprint arXiv:2410.22570},
  year    = {2024},
  doi     = {10.48550/arXiv.2410.22570}
}

@article{Rhodes2025,
  author  = {Rhodes, Benjamin and Vandenhaute, Sander and {\v{S}}imkus, Vaidotas and Gin, James and Godwin, Jonathan and Duignan, Tim and Neumann, Mark},
  title   = {{Orb-v3}: atomistic simulation at scale},
  journal = {arXiv preprint arXiv:2504.06231},
  year    = {2025},
  doi     = {10.48550/arXiv.2504.06231}
}

@article{Park2024,
  author  = {Park, Yutack and Kim, Jaesun and Hwang, Seungwoo and Han, Seungwu},
  title   = {Scalable Parallel Algorithm for Graph Neural Network Interatomic Potentials in Molecular Dynamics Simulations},
  journal = {J. Chem. Theory Comput.},
  volume  = {20},
  number  = {11},
  pages   = {4857--4868},
  year    = {2024},
  doi     = {10.1021/acs.jctc.4c00190}
}

@article{Zhang2019,
  author  = {Zhang, Yubo and Furness, James W. and Xiao, Bing and Sun, Jianwei},
  title   = {Subtlety of {{TiO}$_2$} phase stability: Reliability of the density functional theory predictions and persistence of the self-interaction error},
  journal = {J. Chem. Phys.},
  volume  = {150},
  number  = {1},
  pages   = {014105},
  year    = {2019},
  doi     = {10.1063/1.5055623}
}

@article{Kim2024,
  author  = {Kim, Jaesun and Kim, Jisu and Kim, Jaehoon and Lee, Jiho and Park, Yutack and Kang, Youngho and Han, Seungwu},
  title   = {Data-{Efficient} {Multifidelity} {Training} for {High-Fidelity} {Machine} {Learning} {Interatomic} {Potentials}},
  journal = {J. Am. Chem. Soc.},
  volume  = {147},
  number  = {1},
  pages   = {1042--1054},
  year    = {2024},
  doi     = {10.1021/jacs.4c14455}
}

@article{Bigi2026,
      title={Pushing the limits of unconstrained machine-learned interatomic potentials},
      author={Filippo Bigi and Paolo Pegolo and Arslan Mazitov and Michele Ceriotti},
      year={2026},
      doi = {10.48550/arXiv.2601.16195},
      journal = {arXiv preprint arXiv:2601.16195}
    
}

@inproceedings{Liao2024,
  author    = {Liao, Yi-Lun and Wood, Brandon and Das, Abhishek and Smidt, Tess},
  title     = {{{EquiformerV2}}: {Improved} {Equivariant} {Transformer} for {Scaling} to {Higher-Degree} {Representations}},
  booktitle = {International Conference on Learning Representations (ICLR)},
  year      = {2024},
  url       = {https://openreview.net/forum?id=y90S8YIu6S},
  doi       = {10.48550/arXiv.2306.12059}
}

@inproceedings{Liao2023,
    title={{Equiformer: Equivariant Graph Attention Transformer for 3D Atomistic Graphs}},
    author={Yi-Lun Liao and Tess Smidt},
    booktitle={International Conference on Learning Representations (ICLR)},
    year={2023},
    url={https://openreview.net/forum?id=KwmPfARgOTD}
}

@article{Liao2026,
    title={{EquiformerV3}: {Scaling} {Efficient}, {Expressive}, and {General} {SE}(3)-{Equivariant} {Graph} {Attention} {Transformers}}, 
    author={Yi-Lun Liao and Alexander J. Hoffman and Sabrina C. Shen and Alexandre Duval and Sam Walton Norwood and Tess Smidt},
    journal={arXiv preprint arXiv:2604.09130},
    doi = {10.48550/arXiv.2604.09130},
    year={2026}
}

@article{Chanussot2021,
  author  = {Chanussot, Lowik and Das, Abhishek and Goyal, Siddharth and Lavril, Thibaut and Shuaibi, Muhammed and Riviere, Morgane and Tran, Kevin and Heras-Domingo, Javier and Ho, Caleb and Hu, Weihua and Palizhati, Aini and Sriram, Anuroop and Wood, Brandon and Yoon, Junwoong and Parikh, Devi and Zitnick, C. Lawrence and Ulissi, Zachary},
  title   = {Open {Catalyst} 2020 ({OC20}) {Dataset} and {Community} {Challenges}},
  journal = {ACS Catal.},
  volume  = {11},
  number  = {10},
  pages   = {6059--6072},
  year    = {2021},
  doi     = {10.1021/acscatal.0c04525}
}

@article{Barroso-Luque2024,
  author  = {Barroso-Luque, Luis and Shuaibi, Muhammed and Fu, Xiang and Wood, Brandon M. and Dzamba, Misko and Gao, Meng and Rizvi, Ammar and Zitnick, C. Lawrence and Ulissi, Zachary W.},
  title   = {Open {Materials} 2024 ({OMat24}) {Inorganic} {Materials} {Dataset} and {Models}},
  journal = {arXiv preprint arXiv:2410.12771},
  year    = {2024},
  doi     = {10.48550/arXiv.2410.12771}
}

@article{Fu2025,
  author  = {Fu, Xiang and Wood, Brandon M. and Barroso-Luque, Luis and Levine, Daniel S. and Gao, Meng and Dzamba, Misko and Zitnick, C. Lawrence},
  title   = {Learning {Smooth} and {Expressive} {Interatomic} {Potentials} for {Physical} {Property} {Prediction}},
  journal = {arXiv preprint arXiv:2502.12147},
  year    = {2025},
  doi     = {10.48550/arXiv.2502.12147}
}

@article{Lysogorskiy2025,
  author  = {Lysogorskiy, Yury and Bochkarev, Anton and Drautz, Ralf},
  title   = {Graph atomic cluster expansion for foundational machine learning interatomic potentials},
  journal = {arXiv preprint arXiv:2508.17936},
  year    = {2025},
  doi     = {10.48550/arXiv.2508.17936}
}

@article{Lysogorskiy2024,
  title = {Graph Atomic Cluster Expansion for Semilocal Interactions beyond Equivariant Message Passing},
  author = {Bochkarev, Anton and Lysogorskiy, Yury and Drautz, Ralf},
  journal = {Phys. Rev. X},
  volume = {14},
  issue = {2},
  pages = {021036},
  numpages = {28},
  year = {2024},
  month = {Jun},
  publisher = {American Physical Society},
  doi = {10.1103/PhysRevX.14.021036},
  url = {https://link.aps.org/doi/10.1103/PhysRevX.14.021036}
}

@inproceedings{Passaro2023,
  author    = {Passaro, Saro and Zitnick, C. Lawrence},
  title     = {Reducing {SO}(3) {Convolutions} to {SO}(2) for {Efficient} {Equivariant} {GNNs}},
  booktitle = {Proc. Mach. Learn. Res. (PMLR)},
  volume    = {202},
  pages     = {27424--27438},
  year      = {2023},
  publisher = {PMLR},
  doi       = {10.48550/arXiv.2302.03655}
}

@article{Wedig2013,
  title = {Electronic origin of the structural anomalies of zinc and cadmium},
  author = {Wedig, Ulrich and Nuss, Hanne and Nuss, J{\"{u}}rgen and Jansen, Martin and Andrae, Dirk and Paulus, Beate and Kirfel, Armin and Weyrich, Wolf},
  journal = {Z. Anorg. Allg. Chem.},
  volume = {639},
  number = {11},
  pages = {2036--2046},
  year = {2013},
  doi = {10.1002/zaac.201300091}
}

@article{Hart2021,
  author  = {Hart, Gus L. W. and Mueller, Tim and Toher, Cormac and Curtarolo, Stefano},
  title   = {Machine learning for alloys},
  journal = {Nat. Rev. Mater.},
  volume  = {6},
  pages   = {730--755},
  year    = {2021},
  doi     = {10.1038/s41578-021-00340-w}
}

@article{Mishin2021,
  title = {Machine-learning interatomic potentials for materials science},
  volume = {214},
  ISSN = {1359-6454},
  url = {http://dx.doi.org/10.1016/j.actamat.2021.116980},
  DOI = {10.1016/j.actamat.2021.116980},
  journal = {Acta Materialia},
  publisher = {Elsevier BV},
  author = {Mishin,  Y.},
  year = {2021},
  month = aug,
  pages = {116980}
}

@article{Deringer2019,
  author  = {Deringer, Volker L. and Caro, Miguel A. and Cs{\'a}nyi, G{\'a}bor},
  title   = {Machine Learning Interatomic Potentials as Emerging Tools for Materials Science},
  journal = {Adv. Mater.},
  volume  = {31},
  pages   = {1902765},
  year    = {2019},
  doi     = {10.1002/adma.201902765}
}

@article{Haussermann2001,
  DOI={10.1103/PhysRevB.64.245114},
  title={Origin of the c/a variation in hexagonal close-packed divalent metals},
  author={H{\"a}ussermann, U and Simak, SI},
  journal={Phys. Rev. B},
  volume={64},
  number={24},
  pages={245114},
  year={2001},
  publisher={APS}
}

@article{Pota2024,
  author  = {P{\'o}ta, B. and Ahlawat, P. and Cs{\'a}nyi, G. and Simoncelli, M.},
  title   = {Thermal Conductivity Predictions with Foundation Atomistic Models},
  journal = {arXiv preprint arXiv:2408.00755},
  year    = {2024},
  doi = {10.48550/arXiv.2408.00755}
}

@article{Kresse1994,
  author  = {Kresse, G. and Hafner, J.},
  title   = {{\it Ab initio} molecular-dynamics simulation of the liquid-metal--amorphous-semiconductor transition in germanium},
  journal = {Phys. Rev. B},
  volume  = {49},
  number  = {20},
  pages   = {14251--14269},
  year    = {1994},
  doi     = {10.1103/physrevb.49.14251}
}

@article{Kresse1996B,
  author  = {Kresse, G. and Furthm{\"{u}}ller, J.},
  title   = {Efficiency of  ab-initio total energy calculations for metals and semiconductors using a plane-wave basis set},
  journal = {Comput. Mater. Sci.},
  volume  = {6},
  number  = {1},
  pages   = {15--50},
  year    = {1996},
  doi     = {10.1016/0927-0256(96)00008-0}
}

@article{Kresse1993,
  author  = {Kresse, G. and Hafner, J.},
  title   = {{\it Ab initio} molecular dynamics for liquid metals},
  journal = {Phys. Rev. B},
  volume  = {47},
  number  = {1},
  pages   = {558--561},
  year    = {1993},
  doi     = {10.1103/physrevb.47.558}
}

@article{Langreth1983,
  author  = {Langreth, David C. and Mehl, M. J.},
  title   = {Beyond the local-density approximation in calculations of ground-state electronic properties},
  journal = {Phys. Rev. B},
  volume  = {28},
  number  = {4},
  pages   = {1809--1834},
  year    = {1983},
  doi     = {10.1103/physrevb.28.1809}
}

@article{Perdew1996,
  author  = {Perdew, John P. and Burke, Kieron and Ernzerhof, Matthias},
  title   = {Generalized {Gradient} {Approximation} {Made} {Simple}},
  journal = {Phys. Rev. Lett.},
  volume  = {77},
  number  = {18},
  pages   = {3865--3868},
  year    = {1996},
  doi     = {10.1103/PhysRevLett.77.3865}
}

@article{Furness2020,
  author  = {Furness, James W. and Kaplan, Aaron D. and Ning, Jinliang and Perdew, John P. and Sun, Jianwei},
  title   = {Accurate and Numerically Efficient {{r$^2$SCAN}} Meta-Generalized Gradient Approximation},
  journal = {J. Phys. Chem. Lett.},
  volume  = {11},
  number  = {19},
  pages   = {8208--8215},
  year    = {2020},
  doi     = {10.1021/acs.jpclett.0c02405}
}

@article{Batzner2022,
  author  = {Batzner, Simon and Musaelian, Albert and Sun, Lixin and Geiger, Mario and Mailoa, Jonathan P. and Kornbluth, Mordechai and Molinari, Nicola and Smidt, Tess E. and Kozinsky, Boris},
  title   = {{E(3)}-equivariant graph neural networks for data-efficient and accurate interatomic potentials},
  journal = {Nat. Commun.},
  volume  = {13},
  pages   = {2453},
  year    = {2022},
  doi     = {10.1038/s41467-022-29939-5}
}

@article{Jain2013,
  author  = {Jain, Anubhav and Ong, Shyue Ping and Hautier, Geoffroy and Chen, Wei and Richards, William Davidson and Dacek, Stephen and Cholia, Shreyas and Gunter, Dan and Skinner, David and Ceder, Gerbrand and Persson, Kristin A.},
  title   = {Commentary: The {Materials Project}: A materials genome approach to accelerating materials innovation},
  journal = {APL Mater.},
  volume  = {1},
  pages   = {011002},
  year    = {2013},
  doi     = {10.1063/1.4812323}
}

@article{Pham2023,
  author  = {Pham, Truong-Tho and Nguyen, Duc-Long},
  title   = {First-principles prediction of superconductivity in {Mg}{B}$_3${C}$_3$},
  journal = {Phys. Rev. B},
  year    = {2023},
  volume  = {107},
  number  = {13},
  pages   = {134502},
  doi     = {10.1103/PhysRevB.107.134502}
}

@article{Bartok2010,
  title = {Gaussian Approximation Potentials: The Accuracy of Quantum Mechanics, without the Electrons},
  author = {Bart{\'o}k, Albert P. and Payne, Mike C. and Kondor, Risi and Cs{\'a}nyi, G{\'a}bor},
  journal = {Phys. Rev. Lett.},
  volume = {104},
  pages = {136403},
  year = {2010},
  doi = {10.1103/PhysRevLett.104.136403}
}

@article{ak37,
  author  = {Ibarra-Hern{\'a}ndez, Wilfredo and Hajinazar, Samad and Avenda{\~n}o-Franco, Guillermo and Bautista-Hern{\'a}ndez, Alejandro and Kolmogorov, Aleksey N. and Romero, Aldo H.},
  title   = {Structural search for stable {Mg--Ca} alloys accelerated with a neural network interatomic model},
  journal = {Phys. Chem. Chem. Phys.},
  volume  = {20},
  pages   = {27545--27557},
  year    = {2018},
  doi     = {10.1039/c8cp05314f}
}

@article{Ouyang2015,
  title = {Global minimization of gold clusters by combining neural network potentials and the basin-hopping method},
  volume = {7},
  ISSN = {2040-3372},
  url = {http://dx.doi.org/10.1039/c5nr03903g},
  DOI = {10.1039/c5nr03903g},
  number = {36},
  journal = {Nanoscale},
  publisher = {Royal Society of Chemistry (RSC)},
  author = {Ouyang,  Runhai and Xie,  Yu and Jiang,  De-en},
  year = {2015},
  pages = {14817–14821}
}

@article{Huang2018,
  author  = {Huang, Si-Da and Shang, Cheng and Kang, Pei-Lin and Liu, Zhi-Pan},
  title   = {Atomic structure of boron resolved using machine learning and global sampling},
  journal = {Chem. Sci.},
  volume  = {9},
  pages   = {8644--8655},
  year    = {2018},
  doi     = {10.1039/c8sc03427c}
}

@article{Podryabinkin2019,
  author  = {Podryabinkin, Evgeny V. and Tikhonov, Evgeny V. and Shapeev, Alexander V. and Oganov, Artem R.},
  title   = {Accelerating crystal structure prediction by machine-learning interatomic potentials with active learning},
  journal = {Phys. Rev. B},
  volume  = {99},
  pages   = {064114},
  year    = {2019},
  doi     = {10.1103/PhysRevB.99.064114}
}

@article{Behler2008,
  author  = {Behler, J{\"o}rg and Marto{\v{n}}{\'a}k, Roman and Donadio, Davide and Parrinello, Michele},
  title   = {Metadynamics Simulations of the High-Pressure Phases of Silicon Employing a High-Dimensional Neural Network Potential},
  journal = {Phys. Rev. Lett.},
  volume  = {100},
  pages   = {185501},
  year    = {2008},
  doi     = {10.1103/PhysRevLett.100.185501}
}

@article{Deringer2018,
  author  = {Deringer, Volker L. and Proserpio, Davide M. and Cs{\'a}nyi, G{\'a}bor and Pickard, Chris J.},
  title   = {Data-driven learning and prediction of inorganic crystal structures},
  journal = {Faraday Discuss.},
  volume  = {211},
  pages   = {45--59},
  year    = {2018},
  doi     = {10.1039/c8fd00034d}
}

@article{ak34,
  author  = {Hajinazar, Samad and Shao, Junping and Kolmogorov, Aleksey N.},
  title   = {Stratified construction of neural network based interatomic models for multicomponent materials},
  journal = {Phys. Rev. B},
  volume  = {95},
  pages   = {014114},
  year    = {2017},
  doi     = {10.1103/PhysRevB.95.014114}
}

@article{ak48,
  title   = {{\it Ab initio} study of {{Li-Mg-B}} superconductors},
  author  = {Kafle, Gyanu P. and Tomassetti, Charlsey R. and Mazin, Igor I. and Kolmogorov, Aleksey N. and Margine, Elena R.},
  journal = {Phys. Rev. Mater.},
  volume  = {6},
  number  = {8},
  pages   = {084801},
  year    = {2022},
  month   = {Aug},
  doi     = {10.1103/PhysRevMaterials.6.084801},
  url     = {https://doi.org/10.1103/PhysRevMaterials.6.084801}
}

@article{ak26,
  author  = {Gou, Huiyang and Dubrovinskaia, Natalia and Bykova, Elena and Tsirlin, Alexander A. and Kasinathan, Deepa and Richter, Asta and Merlini, Marco and Hanfland, Michael and Abakumov, Artem M. and Batuk, Dmitry and Dubrovinsky, Leonid},
  title   = {Discovery of a superhard iron tetraboride superconductor},
  journal = {Phys. Rev. Lett.},
  volume  = {111},
  pages   = {157002},
  year    = {2013},
  doi     = {10.1103/PhysRevLett.111.157002}
}

@article{Xiang2024,
  title = {Superconductivity up to 14.2 {K} in {{MnB$_4$}} Under Pressure},
  volume = {37},
  ISSN = {1521-4095},
  url = {http://dx.doi.org/10.1002/adma.202416882},
  number = {4},
  journal = {Advanced Materials},
  publisher = {Wiley},
  author = {Xiang,  Zhe‐Ning and Zhang,  Ying‐Jie and Lu,  Qing and Li,  Qing and Li,  Yiwen and Huang,  Tianheng and Zhu,  Yijie and Ye,  Yongze and Sun,  Jian and Wen,  Hai‐Hu},
  year = {2024},
  month = dec 
}

@article{ak22,
  title   = {Structure, bonding, and possible superhardness of {{CrB$_4$}}},
  author  = {Niu, Haiyang and Wang, Jiaqi and Chen, Xing-Qiu and Li, Dianzhong and Li, Yiyi and Lazar, Petr and Podloucky, Raimund and Kolmogorov, Aleksey N.},
  journal = {Phys. Rev. B},
  volume  = {85},
  number  = {14},
  pages   = {144116},
  year    = {2012},
  month   = {Apr},
  doi     = {10.1103/PhysRevB.85.144116},
  url     = {https://doi.org/10.1103/PhysRevB.85.144116}
}

@article{Knappschneider2013,
  author  = {Knappschneider, Arno and Litterscheid, Christian and Dzivenko, Dmytro and Kurzman, Joshua A. and Seshadri, Ram and Wagner, Norbert and Beck, Johannes and Riedel, Ralf and Albert, Barbara},
  title   = {Possible superhardness of {CrB$_4$}},
  journal = {Inorg. Chem.},
  volume  = {52},
  pages   = {540--542},
  year    = {2013},
  doi     = {10.1021/ic3020404}
}

@article{Knappschneider2014,
  author  = {Knappschneider, Arno and Litterscheid, Christian and George, Nathan C. and Brgoch, Jakoah and Wagner, Norbert and Beck, Johannes and Kurzman, Joshua A. and Seshadri, Ram and Albert, Barbara},
  title   = {{Peierls}-distorted monoclinic {MnB$_4$} with a {Mn--Mn} bond},
  journal = {Angew. Chem. Int. Ed.},
  volume  = {53},
  pages   = {1684--1688},
  year    = {2014},
  doi     = {10.1002/anie.201306548}
}

@article{Gou2014,
  author  = {Gou, Huiyang and Tsirlin, Alexander A. and Bykova, Elena and Abakumov, Artem M. and Van Tendeloo, Gustaaf and Richter, Asta and Ovsyannikov, Sergey V. and Kurnosov, Alexander V. and Trots, Dmytro M. and Kon{\^o}pkov{\'a}, Zuzana and Dubrovinsky, Leonid},
  title   = {{Peierls} distortion, magnetism, and high hardness of manganese tetraboride},
  journal = {Phys. Rev. B},
  volume  = {83},
  pages   = {064108},
  year    = {2014},
  doi     = {10.1103/PhysRevB.89.064108}
}

@article{Wang1978,
  author  = {Wang, F. E. and Mitchell, M. A. and Sutula, R. A. and Holden, J. R. and Bennett, L. H.},
  title   = {Crystal structure study of a new compound {{Li}$_5${B}$_4$}},
  journal = {J. Less-Common Met.},
  volume  = {61},
  number  = {2},
  pages   = {237--251},
  year    = {1978},
  doi     = {10.1016/0022-5088(78)90219-9}
}

@article{Worle2000,
  title = {Infinite, linear, unbranched borynide chains in {{LiB}$_x$}—{Isoelectronic} to polyyne and polycumulene},
  author = {W{\"{o}}rle, Michael and Nesper, Reinhard},
  journal = {Angew. Chem.},
  volume = {112},
  number = {13},
  pages = {2439--2443},
  year = {2000},
  doi = {10.1002/1521-3757(20000703)112:13<2439::AID-ANGE2439>3.0.CO;2-Q}
}

@article{ak09,
  title = {Theoretical study of metal borides stability},
  author = {Kolmogorov, Aleksey N. and Curtarolo, Stefano},
  journal = {Phys. Rev. B},
  volume = {74},
  pages = {224507},
  year = {2006},
  doi = {10.1103/PhysRevB.74.224507}
}

@article{ak30,
  author  = {Kolmogorov, A. N. and Hajinazar, S. and Angyal, C. and Kuznetsov, V. L. and Jephcoat, A. P.},
  title   = {Synthesis of a predicted layered {{LiB}} via cold compression},
  journal = {Phys. Rev. B},
  volume  = {92},
  number  = {14},
  pages   = {144110},
  year    = {2015},
  doi     = {10.1103/PhysRevB.92.144110}
}

@article{ak13,
  title = {{\it Ab initio} modeling of {Li-B-H} boron-chain alloys for hydrogen storage applications},
  author = {Kolmogorov, A. N. and Drautz, R. and Pettifor, D. G.},
  journal = {Phys. Rev. B},
  volume = {76},
  pages = {184102},
  year = {2007},
  doi = {10.1103/PhysRevB.76.184102}
}

@article{Hermann2012,
  title = {{LiB} and its boron-deficient variants under pressure},
  author = {Hermann, Andreas and Suarez-Alcubilla, Ainhoa and Gurtubay, Idoia G. and Yang, Li-Ming and Bergara, Aitor and Ashcroft, Neil W. and Hoffmann, Roald},
  journal = {Phys. Rev. B},
  volume = {86},
  pages = {144110},
  year = {2012},
  doi = {10.1103/PhysRevB.86.144110}
}

@article{Gubaev2019,
  author  = {Gubaev, Konstantin and Podryabinkin, Evgeny V. and Hart, Gus L. W. and Shapeev, Alexander V.},
  title   = {Accelerating high-throughput searches for new alloys with active learning of interatomic potentials},
  journal = {Comput. Mater. Sci.},
  volume  = {156},
  pages   = {148--156},
  year    = {2019},
  doi     = {10.1016/j.commatsci.2018.09.031}
}

@article{Artrith2011,
  author  = {Artrith, Nongnuch and Morawietz, Tobias and Behler, J{\"o}rg},
  title   = {High-dimensional neural-network potentials for multicomponent systems: Applications to zinc oxide},
  journal = {Phys. Rev. B},
  volume  = {83},
  pages   = {153101},
  year    = {2011},
  doi     = {10.1103/PhysRevB.83.153101}
}

@article{Dolgirev2016,
  author  = {Dolgirev, Pavel E. and Kruglov, Ivan A. and Oganov, Artem R.},
  title   = {Machine learning scheme for fast extraction of chemically interpretable interatomic potentials},
  journal = {AIP Adv.},
  volume  = {6},
  pages   = {085318},
  year    = {2016},
  doi     = {10.1063/1.4961886}
}

@article{Blochl1994B,
  author  = {Bl{\"{o}}chl, Peter E. and Jepsen, O. and Andersen, O. K.},
  title   = {Improved tetrahedron method for {Brillouin}-zone integrations},
  journal = {Phys. Rev. B},
  volume  = {49},
  number  = {23},
  pages   = {16223--16233},
  year    = {1994},
  doi     = {10.1103/PhysRevB.49.16223}
}

@article{Mazitov2025,
  author  = {Mazitov, Arslan and Bigi, Filippo and Kellner, Matthias and Pegolo, Paolo and Tisi, Davide and Fraux, Guillaume and Pozdnyakov, Sergey and Loche, Philip and Ceriotti, Michele},
  title   = {{PET-MAD} as a lightweight universal interatomic potential for advanced materials modeling},
  journal = {Nat. Commun.},
  volume  = {16},
  pages   = {65662},
  year    = {2025},
  doi     = {10.1038/s41467-025-65662-7}
}

@article{Sharma2025,
  author  = {Sharma, Kartikeya and Loew, Antoine and Wang, Haiyuan and Nilsson, Fredrik A. and Jain, Manjari and Marques, Miguel A. L. and Thygesen, Kristian S.},
  title   = {Accelerating point defect photo-emission calculations with machine learning interatomic potentials},
  journal = {npj Comput. Mater.},
  volume  = {11},
  pages   = {18201},
  year    = {2025},
  doi     = {10.1038/s41524-025-01820-1}
}

@article{Klime2009,
  author  = {Klime{\v{s}}, Jiri and Bowler, David R. and Michaelides, Angelos},
  title   = {Chemical accuracy for the {van der Waals} density functional},
  journal = {J. Phys.: Condens. Matter},
  volume  = {22},
  pages   = {022201},
  year    = {2009},
  doi     = {10.1088/0953-8984/22/2/022201}
}

@article{Ning2022,
  author  = {Ning, Jinliang and Kothakonda, Manish and Furness, James W. and Kaplan, Aaron D. and Ehlert, Sebastian and Brandenburg, Jan Gerit and Perdew, John P. and Sun, Jianwei},
  title   = {Workhorse minimally empirical dispersion-corrected density functional with tests for weakly bound systems: {{r$^2$SCAN}} + {{rVV10}}},
  journal = {Phys. Rev. B},
  volume  = {106},
  pages   = {075422},
  year    = {2022},
  doi     = {10.1103/PhysRevB.106.075422}
}

@article{Chakraborty2020,
  author  = {Chakraborty, D. and Berland, K. and Thonhauser, T.},
  title   = {Next-{Generation} {Nonlocal} {van der Waals} {Density} {Functional}},
  journal = {J. Chem. Theory Comput.},
  volume  = {16},
  number  = {9},
  pages   = {5893--5911},
  year    = {2020},
  doi     = {10.1021/acs.jctc.0c00471}
}

@article{ak56,
  author  = {Gochitashvili, Daviti and Meyers, Maxwell and Wang, Cindy and Kolmogorov, Aleksey N.},
  title   = {Improving structure search with hyperspatial optimization and {TETRIS} seeding},
  journal = {Phys. Chem. Chem. Phys.},
  volume  = {27},
  number  = {47},
  pages   = {25636--25647},
  year    = {2025},
  doi     = {10.1039/D5CP02412A}
}

@article{Takemura2019,
  author  = {Takemura, Kenichi},
  title   = {The zinc story under high pressure},
  journal = {J. Miner. Mater. Charact. Eng.},
  volume  = {7},
  number  = {5},
  pages   = {354--372},
  year    = {2019},
  doi     = {10.4236/jmmce.2019.75024}
}

@article{Qu2024,
  title = {Leveraging language representation for materials exploration and discovery},
  volume = {10},
  ISSN = {2057-3960},
  url = {http://dx.doi.org/10.1038/s41524-024-01231-8},
  DOI = {10.1038/s41524-024-01231-8},
  number = {1},
  journal = {npj Comput. Mater.},
  publisher = {Springer Science and Business Media LLC},
  author = {Qu,  Jiaxing and Xie,  Yuxuan Richard and Ciesielski,  Kamil M. and Porter,  Claire E. and Toberer,  Eric S. and Ertekin,  Elif},
  year = {2024},
  month = Mar 
}

@article{Fasoulis2024,
  title = {{RankMHC}: {Learning} to {Rank} {Class-I} {Peptide-MHC} {Structural} {Models}},
  volume = {64},
  ISSN = {1549-960X},
  url = {http://dx.doi.org/10.1021/acs.jcim.4c01278},
  DOI = {10.1021/acs.jcim.4c01278},
  number = {23},
  journal = {Journal of Chemical Information and Modeling},
  publisher = {American Chemical Society (ACS)},
  author = {Fasoulis,  Romanos and Paliouras,  Georgios and Kavraki,  Lydia E.},
  year = {2024},
  month = Nov,
  pages = {8729–8742}
}

@article{Ahn2016,
  title = {Pressure-structure relationships in the 10 {K} layered carbide halide superconductor {Y$_2$C$_2$I$_2$}},
  volume = {28},
  ISSN = {1361-648X},
  url = {http://dx.doi.org/10.1088/0953-8984/28/37/375703},
  DOI = {10.1088/0953-8984/28/37/375703},
  number = {37},
  journal = {Journal of Physics: Condensed Matter},
  publisher = {IOP Publishing},
  author = {Ahn,  Kyungsoo and Kremer,  Reinhard K and Simon,  Arndt and Marshall,  William G and Muñoz,  Alfonso},
  year = {2016},
  month = Jul,
  pages = {375703}
}

@article{Henn1996,
  title = {{Bulk} {Superconductivity} at 10 {K} in the {Layered} {Compounds} {Y$_2$C$_2$I$_2$} and {Y$_2$C$_2$Br$_2$}},
  volume = {77},
  ISSN = {1079-7114},
  url = {http://dx.doi.org/10.1103/PhysRevLett.77.374},
  DOI = {10.1103/physrevlett.77.374},
  number = {2},
  journal = {Phys. Rev. Lett.},
  publisher = {American Physical Society (APS)},
  author = {Henn,  R. W. and Schnelle,  W. and Kremer,  R. K. and Simon,  A.},
  year = {1996},
  month = Jul,
  pages = {374–377}
}

@article{Berthold1976,
  title = {Die {Kristallstruktur} des {AgClO$_4$}},
  volume = {144},
  ISSN = {2194-4946},
  url = {http://dx.doi.org/10.1524/zkri.1976.144.16.116},
  DOI = {10.1524/zkri.1976.144.16.116},
  number = {1-6},
  journal = {Zeitschrift f\"{u}r Kristallographie - Crystalline Materials},
  publisher = {Walter de Gruyter GmbH},
  author = {Berthold,  H. J. and Molepo, J. M. and Wartchow, R.},
  year = {1976},
  month = Dec,
  pages = {116–125}
}

@article{Hlukhyy2004,
  title = {{The} {Hexagonal} {Laves} {Phase} {MgIr$_2$}},
  volume = {59},
  ISSN = {0932-0776},
  url = {http://dx.doi.org/10.1515/znb-2004-0813},
  DOI = {10.1515/znb-2004-0813},
  number = {8},
  journal = {Zeitschrift f\"{u}r Naturforschung B},
  publisher = {Walter de Gruyter GmbH},
  author = {Hlukhyy,  Viktor and P\"{o}ttgen,  Rainer},
  year = {2004},
  month = Aug,
  pages = {943–946}
}

@article{Becker2000,
  author  = {Becker, M. and Nuss, J. and Jansen, M.},
  title   = {{Synthese} und {Charakterisierung} von {Natriumcyanamid}},
  journal = {Zeitschrift f{\"u}r anorganische und allgemeine Chemie},
  year    = {2000},
  volume  = {626},
  number  = {12},
  pages   = {2505--2508},
  doi     = {10.1002/1521-3749(200012)626:12<2505::AID-ZAAC2505>3.0.CO;2-\#}
}

\end{document}